\documentclass{article}

\usepackage{PRIMEarxiv}

\usepackage[utf8]{inputenc} 
\usepackage[T1]{fontenc}    
\usepackage{hyperref}       
\usepackage{url}            
\usepackage{booktabs}       
\usepackage{amsfonts}       
\usepackage{amsmath}
\usepackage{amssymb,amsthm,mathtools}
\usepackage{nicefrac}       
\usepackage{microtype}      
\usepackage{lipsum}
\usepackage{fancyhdr}       
\usepackage{graphicx}       
\usepackage{natbib}
\graphicspath{{media/}}     

\title{Perceptogram: Reconstructing Visual Percepts and Presumptive Electrode Preference from EEG
}

\author{
  Teng Fei, Srinivas Ravishankar, Zhining Chen, Abhinav Uppal, Ian Jackson, Virginia R. de Sa \\
  Cognitive Science \\
  University of California, San Diego \\
  La Jolla\\
  \texttt{\{tfei, srravishankar, zhc008, auppal, ijackson\}@ucsd.edu} \\
}

\begin{document}
\maketitle

\begin{abstract}
Visual neural decoding from EEG has improved significantly due to diffusion models that can reconstruct high-quality images from decoded latents. 
While recent works have focused on relatively complex architectures to achieve good reconstruction performance from EEG, less attention has been paid to the source of this information. 
We present a unified framework that not only enables image reconstruction from EEG using a simple linear decoder, but also isolates interpretable EEG feature maps that support visual perception.
Unlike prior approaches that rely on deep, opaque models, our method leverages the inherent structure of CLIP embeddings to keep the mapping linear.
We show that training a simple linear decoder from EEG to CLIP latent space, followed by a frozen pre-trained diffusion model, is sufficient to decode images with state-of-the-art reconstruction performance.
Beyond reconstruction, Perceptogram enables the visualization of presumptive electrode preference and EEG patterns, revealing interpretable EEG feature maps that correspond to distinct visual attributes, such as semantic class, texture, and hue.
We thus use our framework, Perceptogram, to probe EEG signals at various levels of the visual information hierarchy. 
We make our code publicly available: \href{https://github.com/desa-lab/Perceptogram}{https://github.com/desa-lab/Perceptogram}
\end{abstract}

\keywords{EEG \and Visual Reconstruction \and Brain-Computer Interface \and Representational Alignment}

\section{Introduction}

Scalp EEG’s low signal-to-noise ratio \cite{low_snr} and spatial resolution \cite{Burle2015} has long constrained its application to basic brain-state decoding, such as motor imagery, attention, and surprisal responses (e.g. P300).
While recent advances have extended EEG’s utility to visual decoding, these approaches predominantly rely on deep, non-linear architectures \cite{spampinato2016, Li_Wei_Li_Zou_Liu_2024} that, while effective, often obscure the underlying neural representations. This reliance on multi-layered networks has solidified the belief that meaningful visual decoding from EEG demands complex non-linear transformations, prioritizing performance at the cost of interpretability.

We challenge this assumption with Perceptogram -- a linear framework that redefines EEG-based visual reconstruction by leveraging the inherent structure of CLIP embeddings through a linear mapping from EEG signals directly into CLIP’s latent space. These latents are then sent to a frozen, pre-trained diffusion model, yielding high-fidelity visual reconstructions. By eliminating the need for deep non-linear stacks, Perceptogram reveals that CLIP’s latent space is intrinsically compatible with the representational dynamics of scalp EEG, enabling accurate and interpretable decoding.

\subsection{Related Work}
fMRI studies of visual \citep{Kay_Naselaris_Prenger_Gallant_2008} and semantic representations \citep{mitchell_predicting_2008} laid the groundwork for recent impressive visual reconstructions  \citep{takagi_high-resolution_2023}. Earlier EEG reconstructions suffered from poor experiment design \citep{Li_Johansen_Ahmed_Ilyevsky_Wilbur_Bharadwaj_Siskind_2020} which, combined with inadequate quality control in the data, led to inflated reconstruction results due to classifying noise statistics \citep{spampinato2016}. The THINGS-EEG2 dataset \citep{giffordLargeRichEEG2022}, with improvements in experiment design and quality control, prompted a resurgence of EEG visual decoding \cite{nice_eeg2024, li2024}. These subsequent studies developed more complicated decoding pipelines
but did not address the organizing principles of the underlying EEG features.

The advent of vision-language foundation models such as CLIP, trained on very large (400M images) datasets, has produced rich latent space representations that capture the high-level semantic structure between images \cite{clipvision}.
These models enable more efficient encoding of visual-semantic relationships, which has been recently exploited in fMRI-based decoding \cite{Ozcelik_VanRullen_2023}. Motivated by these findings, we hypothesize that if CLIP embeddings and EEG representations share high-level relational structure, then a simple linear mapping could bridge these spaces for high-performance image reconstruction--circumventing the need for deep, multi-layered networks. This premise forms the foundation of our reconstruction pipeline.

\begin{figure}[t]
    \centering
    \includegraphics[width=0.5\columnwidth]{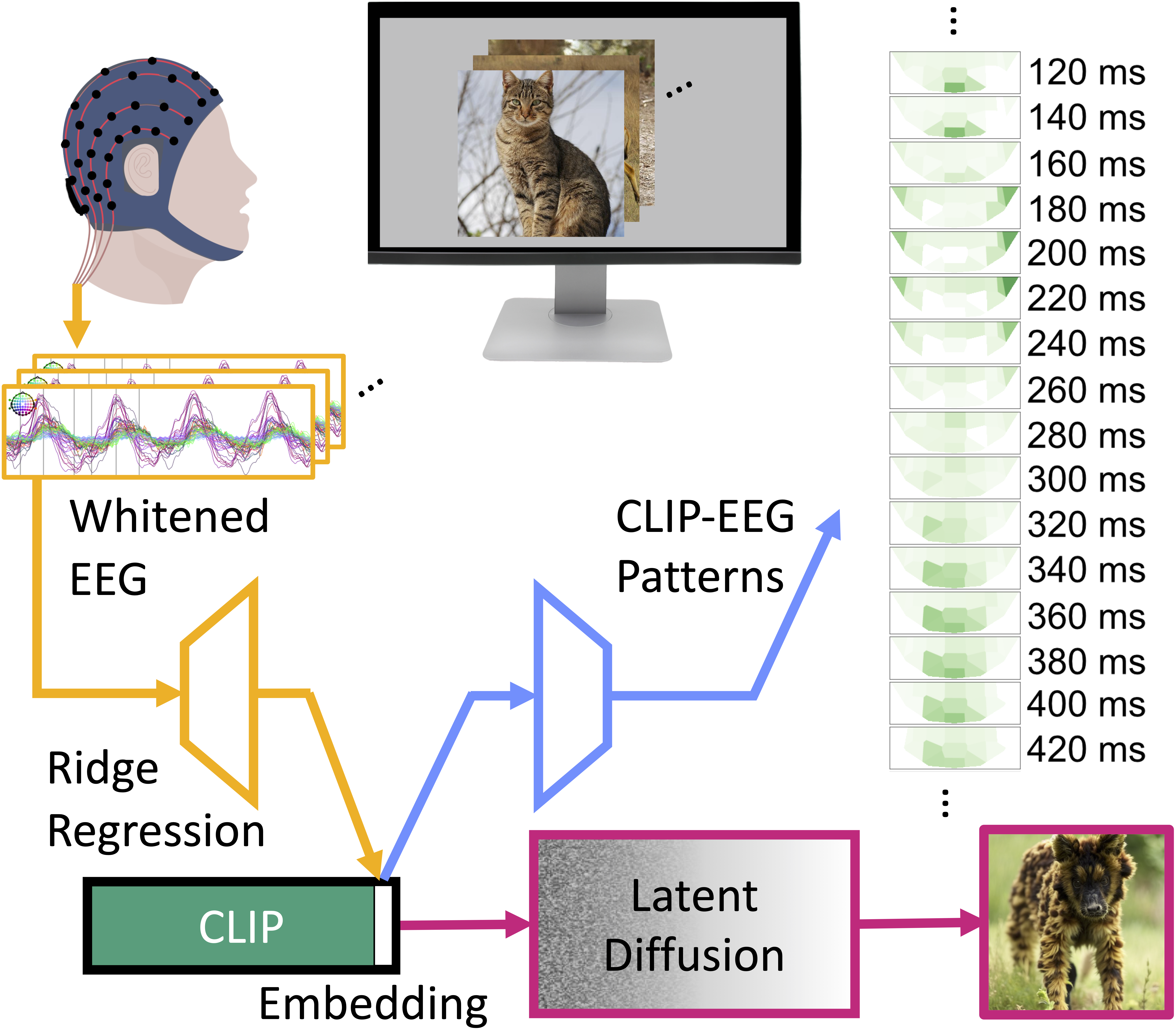}%
    \caption{Pipeline overview: There are three primary components: A linear \textit{decoder} (orange) from brain space to latent space, a linear \textit{encoder} (blue) 
    mapping this decoded latent back into brain space, and a \textit{reconstructer} (purple) that generates an image from the decoder output. The encoder output is a latent-filtered spatio-temporal brain pattern for that image.
    }
    \label{fig:overview}
\end{figure}

\paragraph{Our Approach}

Our approach goes beyond mere image reconstruction from EEG–it serves as a tool to
isolate and interpret the EEG features that support the decoding. We do this in two ways: 
{\bf 1. Test-time Perturbation:} we systematically perturb EEG inputs (mirroring electrodes, swapping time segments) during test time, and inspect how this changes the original reconstructions.
{\bf 2. Decoding-Encoding Loop:} At the heart of our method is a linear mapping from EEG to CLIP’s latent space. 
Once decoded, the predicted latents are projected back to EEG space using a linear encoder. 
Intuitively, this decoding-encoding loop, wherein EEG is decoded to latent space and then encoded back, acts as a filter to isolate EEG features (electrodes and time points) carrying shared semantic information.
The result is what we term latent-filtered (spatiotemporal) EEG patterns, directly analogous to common spatial patterns \citep{Blankertz2008} in the Brain-Computer Interface (BCI) literature.
To explore the full spectrum of visual processing, we extend this analysis beyond high-level concepts to low-level visual features like color and texture.
By replacing CLIP with latent spaces emphasizing these features, we isolate the corresponding spatiotemporal EEG components, uncovering how the brain represents these lower-level visual features.
We extend our approach with cross-modal validation using fMRI data from a similar experimental paradigm allowing us to verify the spatial localization  of latent-filtered EEG patterns.
Together, these components form a unified pipeline that enables both reconstruction and neuroscientific insight into the brain’s dynamic representation of visual categories.

\section{Methods}

\subsection{Dataset}
We used the publicly available THINGS-EEG2 \citep{Gifford_Dwivedi_Roig_Cichy_2022} and Natural Scenes Dataset (NSD) \citep{allen2022massive} for EEG and fMRI analyses, respectively, to validate findings from our EEG analysis. Both datasets are described in the Appendix \ref{appendix:dataset} and briefly introduced below.

\subsubsection{THINGS-EEG2}
EEG data was collected from 10 subjects viewing a set of 16740 images including 200 test images, each presented for 100 ms with 100 ms inter-trial interval. 
Each training image was shown 4 times, whereas each test image was shown 80 times, in pseudo-randomized presentation order. Preprocessed data obtained from \url{https://osf.io/anp5v/} consisted of 17 posterior EEG channels (of 63 total), down-sampled from 1000 Hz to 100 Hz. Trials of 0.8 second duration (80 samples at 100 Hz) were extracted relative to stimulus onset, and averaged across image repetitions within subjects. 80 samples times 17 channels or 1360 dimensional trials were thus obtained per image per subject. As images were presented every 200ms, effects of random subsequent images on individual trial responses were mitigated by averaging over trials.

\subsubsection{NSD}
7T fMRI data was collected for Microsoft's COCO images database \citep{coco_dataset_2014} with images presented for 3 seconds and 1 second inter-trial interval. 982 images shared across 4 trial-complete subjects were used as the test set, while each subject had 8859 exclusive images (not shared across subjects) that were used as the training set. Image presentation order was pseudo-randomized across the entire image set, with each image presented 3 times to enhance the signal-to-noise ratio. The original dataset includes the preprocessed betas commonly used in decoding, which is what we used.

\begin{figure*}[h] 
    \centering    
    \includegraphics[width=0.8\linewidth]{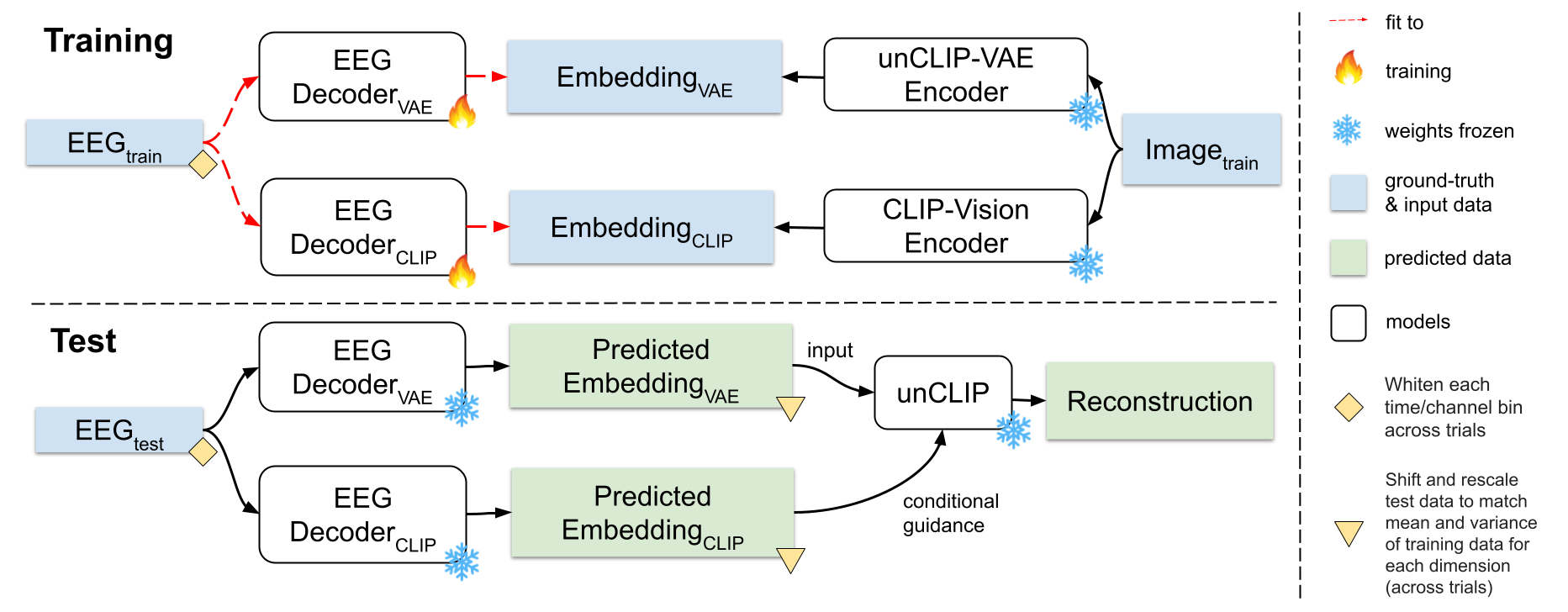}
    \caption{Flowchart illustrating image reconstruction using CLIP as latent space, and unCLIP as reconstructer.  During the Test stage, the test EEG is fed though the 2 matrices to get the predicted VAE and CLIP-Vision latents. unCLIP then turns the predicted VAE and CLIP-Vision latents into actual images. }
    \label{fig:flowchart_unCLIP}
\end{figure*}

\subsection{Model Architecture}
As shown in Fig. \ref{fig:overview}, there are three primary components in our pipeline: A linear \textit{decoder} from brain space to latent space, a linear \textit{encoder} mapping this decoded latent back into brain space, and a \textit{reconstructer} that generates an image from the decoder's output latent. The input to the pipeline is brain data, and two outputs are produced. These correspond to the reconstructed image and the latent-filtered brain patterns for that image.

\subsubsection{Encoder, Decoder}
We employ linear regression in both the encoder and decoder, shown to be an effective way to decode latents from fMRI \citep{Ozcelik_VanRullen_2023}. Regularization strength ($\alpha$) is chosen based on the correlation of Subject-1's predicted latent and the ground truth latent (Table. \ref{tab:latent_reg_strengths} has all $\alpha$'s).

\begin{table}[h]
    \centering
    \caption{Regularization strengths ($\alpha$) for each latent representation, modality, and mapping direction.}
    \label{tab:latent_reg_strengths}
    \begin{tabular}{@{}lll@{}}
        \toprule
        Latent & Mapping (modality–direction) & $\alpha$ \\
        \midrule
        unCLIP CLIP                        & EEG decode  & 0 \\
        unCLIP CLIP                        & EEG encode  & 1 000 \\
        unCLIP CLIP Grayscale              & EEG decode  & 1 000 \\
        unCLIP CLIP Grayscale              & EEG encode  & 1 000 \\
        unCLIP VAE                         & EEG decode  & 0 \\
        PCA                                & EEG decode  & 1 000 \\
        PCA                                & EEG encode  & 1 000 000 \\
        ICA                                & EEG decode  & 1 000 \\
        ICA                                & EEG encode  & 1 000 000 \\
        VDVAE                              & EEG decode  & 1 000 \\
        VDVAE                              & EEG encode  & 1 000 \\
        Versatile Diffusion CLIP-Vision    & EEG decode  & 1 000 \\
        Versatile Diffusion CLIP-Text      & EEG decode  & 10 000 \\
        \midrule
        unCLIP VAE                         & fMRI decode & 100 000 \\
        unCLIP CLIP                        & fMRI decode & 100 000 \\
        unCLIP CLIP                        & fMRI encode & 10 000 \\
        PCA                                & fMRI decode & 100 000 \\
        PCA                                & fMRI encode & 100 000 \\
        ICA                                & fMRI decode & 100 000 \\
        ICA                                & fMRI encode & 100 000 \\
        VDVAE                              & fMRI decode & 100 000 \\
        VDVAE                              & fMRI encode & 1 000 000 \\
        Versatile Diffusion CLIP-Vision    & fMRI decode & 50 000 \\
        Versatile Diffusion CLIP-Text      & fMRI decode & 100 000 \\
        \midrule
        unCLIP CLIP                        & MEG decode  & 10 000 \\
        unCLIP VAE                         & MEG decode  & 10 000 \\
        \bottomrule
    \end{tabular}
\end{table}

\subsubsection{Latent Space}
While Fig. \ref{fig:overview} illustrates the pipeline using the CLIP latent space, we also use other latent spaces which emphasize different visual features in their reconstructions. With CLIP latents, the reconstructions preserve high-level semantic categories of the images. Latents from VDVAE, PCA or ICA might emphasize other lower-level visual features. 

\subsubsection{Reconstructor} 
Reconstructions from CLIP latents require a diffusion module. We have used both Versatile Diffusion, as in \cite{Ozcelik_VanRullen_2023}, and unCLIP \citep{ramesh_hierarchical_2022}. Images are reconstructed from VDVAE latents using its pretrained frozen decoder, and PCA/ICA simply use linear inverse projections. 

\subsubsection{Implementation details}
Linear models were trained on an AMD EPYC 7302 16-Core Processor using the scikit-learn framework, with each training run completing in a few seconds. Reconstructions using pretrained latent diffusion models were executed on a single NVIDIA A6000 GPU, with an average runtime of approximately 1.5 seconds per image.

\subsection{Evaluation}

\subsubsection{Reconstruction}

We used the same performance metrics 
(see Table \ref{tab:performance}) as in \cite{Ozcelik_VanRullen_2023}, which has been used in other followup studies such as MindEye \cite{Scotti_Banerjee_Goode_Shabalin_Nguyen_Cohen_Dempster_Verlinde_Yundler_Weisberg_et_al_2023}. The 8 metrics we used are Pixel Correlation (PixCorr), Structural Similarity (SSIM), AlexNet layer 2 and 5 outputs pairwise correlations, InceptionNet output pairwise correlation, CLIP ViT output pairwise correlation, EfficientNet output distance, and SwAV output distance.
PixCorr and SSIM involve comparing the reconstructed image with the ground-truth (GT) test image. PixCorr is a low-level (pixel) measure that involves vectorizing the reconstructed and GT images and computing the correlation coefficient between the resulting vectors. SSIM is a measure developed by Wang et al. 2004 that computes a match value between GT and reconstructed images as a function of overall luminance match, overall contrast match, and a ``structural'' match which is defined by normalizing each image by its mean and standard deviation. 

\subsubsection{Retrieval}
To assess the alignment between projected EEG representations and CLIP embeddings, we employed a cross-modal retrieval paradigm. Specifically, we measured Recall@1 and Recall@5, standard retrieval metrics quantifying how often the ground-truth target embedding appears among the top-k nearest neighbors in the retrieval set. For each projected EEG sample, we computed l-2 normalized cosine similarity with all CLIP embeddings in the test set and ranked them by similarity score. Recall@1 reflects the proportion of trials where the correct CLIP embedding was retrieved as the most similar item, while Recall@5 captures retrieval within the top five matches.
\subsection{Experiments}

\subsubsection{Image reconstruction}
We started by using the Versatile Diffusion method of \cite{Ozcelik_VanRullen_2023} but later developed a simpler pipeline using unCLIP \citep{ramesh_hierarchical_2022} which we use for most of our qualitative analyses as it simplifies the overall architecture while maintaining comparable reconstruction performance. A flowchart illustrating image reconstruction for the unCLIP variant of our pipeline is shown in Fig. \ref{fig:flowchart_unCLIP}.  We introduce an initial VAE encoder into the standard unCLIP framework so our unCLIP diffusion process uses two types of latents as input, an initial (VAE) latent and a conditioning/guidance vector (CLIP).

In the training stage, an image is processed through unCLIP-VAE and CLIP-Vision encoders, producing two target latent embeddings needed for our unCLIP diffuser. Linear decoders are trained using linear regression to map EEG to these targets. Before fitting, the EEG data is whitened to normalize each of the 1360 EEG dimensions across the 16540 training classes. In the test stage, the decoder predicts corresponding embeddings given an EEG trial. The predicted embeddings are shifted and rescaled using the mean and variance of the training latent distribution. These are input to the unCLIP decoder to produce a reconstructed image. All pre-trained models are frozen, and the linear decoders are the only modules trained. The Versatile Diffusion variant works similarly and is illustrated in the appendix (Fig. \ref{fig:flowchart_VD}).

\subsubsection{Electrode preference visualization by visually reconstructing the encoding weights}
To investigate the semantic tuning properties in our neural-to-CLIP mapping, we performed a detailed analysis of the encoding weights that transform CLIP latent representations into predicted EEG waveforms. These weights provide insight into what visual features maximally drive neural responses at each electrode and time point.

We extracted individual weight vectors from the CLIP-to-EEG encoding matrix, where each vector corresponds to a specific electrode at a given time bin. Each weight vector has the same dimensionality as CLIP embeddings (1024 dimensions) and represents the linear transformation from CLIP space to predicted EEG voltage. To visualize the preferred stimulus features encoded by these weights, we treated each weight vector as if it were a CLIP embedding of an actual image and reconstruct it using unCLIP. This approach reveals the "ideal stimulus" that would maximally activate each electrode according to the learned linear model.

\subsubsection{Electrode mirroring}
Motivated by the well known contralateral organization of visual cortex -- where stimuli in the right visual field are initially processed by the left visual cortex, and those in the left field by the right cortex -- we investigated if we map our EEG to the homologous electrodes across the midline, whether we could see the reconstructed image mirror about the vertical midline.
We train the model using unaltered EEG as described in the previous section. During evaluation, we examine reconstructions produced by feeding EEG data mirrored along the midline. For example, during test time, we would swap the EEG between channels O1 and O2 (see Fig. \ref{fig:topography} for the electrode topography). We consider two mirrored conditions: In one, we swap the EEG of all non-midline electrodes, and in the other, we swap only EEG for O1 and O2 (over primary visual cortex). 

\subsubsection{Time-Swapping}
In order to investigate the temporal dynamics and the salient features in the EEG data, we develop a novel technique to find time-ranges that are most sensitive to disturbance. We used pairs of images and swapped analogous time segments of data between EEG responses to each of the images as demonstrated in Fig. \ref{fig:appendix_timeswap}.

Each image results from reconstruction of EEG where a 120ms time window centered at the corresponding time point is swapped between the 2 classes within that window while holding the signal outside the window the same. On top of each reconstructed image, we added a color bar that proportionally indicates which EEG time segment is swapped with the other class for that image. The two classes are represented by red and blue in this color bar and time is represented in the horizontal direction so a blue bar with a small red square represents that 120ms of the EEG at its relative location is swapped with the EEG for the other class. The small squares progress to the right as the samples progress to the right. The original, unswapped reconstructions (shown at right) have their color bars all blue/red, indicating that no part of their EEG is swapped with the other class.

\subsubsection{Latent-filtered EEG Patterns}
\begin{figure}
    \centering
    \includegraphics[width=0.45\linewidth]{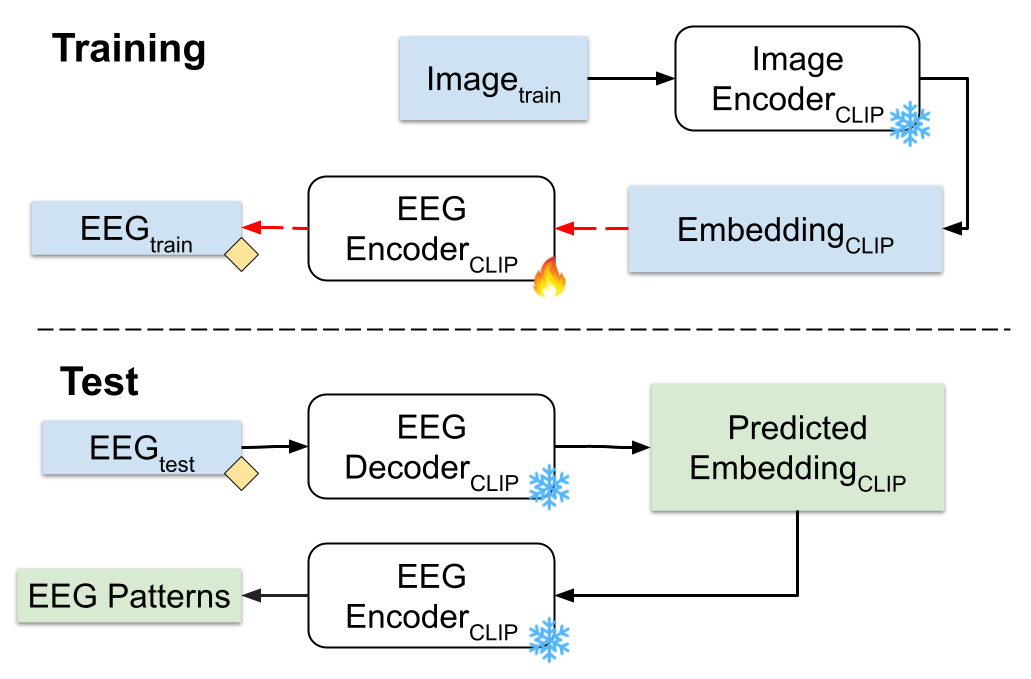}
    \caption{Flowchart illustrating how to produce the EEG patterns linked to the CLIP embedding. It is similar to the regular unCLIP pipeline (Fig. \ref{fig:flowchart_unCLIP}). The main difference here is that we train an encoding model predicting EEG from CLIP.}
    \label{fig:clip_pattern_flowchart}
\end{figure}

This section describes how we obtain the spatio-temporal EEG patterns specific to a visual feature, with textured vs smooth patterns as an example. 
First, we choose a latent space emphasizing the feature of interest, by manually inspecting reconstructions from various latent spaces. Textures appear to be emphasized in reconstructions from VDVAE, as seen in Fig \ref{fig:low-level}.
The encoder and decoder are then trained as described previously. During testing, the decoder is used to first predict image latents from the held-out EEG data, and the encoder is used to project these latents back to EEG space, thus `filtering' the EEG through the chosen latent space. This is illustrated in Fig. \ref{fig:clip_pattern_flowchart}, which highlights the decoding-encoding loop during testing. 

The decoder-encoder loop projects the EEG onto the cross-covariance subspace shared between the EEG and the chosen latent space, so only the EEG features covarying with the chosen latent space would be preserved. In an extreme case if the latent space is uncorrelated with the EEG, then the projected EEG will reveal no structure:

\newcommand{\col}{\operatorname{col}}
\newcommand{\R}{\mathbb{R}}
\newtheorem{lemma}{Lemma}


\paragraph{Setup.}
Let $X\in\R^{n\times d}$ denote EEG features (rows = $n$ classes, columns = $d$ dimensions)
and $Y\in\R^{n\times k}$ denote latent embeddings (columns = $k$ dimensions).
Define ridge-regularized decoder and encoder
\[
W_d := (X^\top X+\lambda_d I_d)^{-1}X^\top Y,\qquad
W_e := (Y^\top Y+\lambda_e I_k)^{-1}Y^\top X .
\]
Let the composite filter be
\[
F := W_e^\top W_d^\top .
\]
Introduce the shorthand
\begin{equation}
\Sigma_{xx} := X^\top X+\lambda_d I_d,\quad
\Sigma_{yy} := Y^\top Y+\lambda_e I_k,\quad
\Sigma_{xy} := X^\top Y,\quad
\Sigma_{yx} := Y^\top X=\Sigma_{xy}^\top .
\label{eq:sigmas}
\end{equation}

\begin{lemma}[Column space of the composite filter]\label{lem:colspace}
With the definitions above,
\[
\col(F)=\col(\Sigma_{xy}).
\]
\end{lemma}

\begin{proof}
First note
\begin{align}
F
&= W_e^\top W_d^\top
= (\Sigma_{yy}^{-1}\Sigma_{yx})^\top(\Sigma_{xx}^{-1}\Sigma_{xy})^\top \notag\\
&= \Sigma_{xy}\,\Sigma_{yy}^{-1}\,\Sigma_{yx}\,\Sigma_{xx}^{-1}
= \Sigma_{xy}\bigl(\Sigma_{yy}^{-1}\Sigma_{yx}\Sigma_{xx}^{-1}\bigr)
= \Sigma_{xy}\,C,
\label{eq:Ffactor}
\end{align}
where
\begin{equation}
C:=\Sigma_{yy}^{-1}\Sigma_{yx}\Sigma_{xx}^{-1}.
\label{eq:Cdef}
\end{equation}
By the basic fact $\col(AB)\subseteq\col(A)$ for conformable $A,B$, we get
\begin{equation}
\col(F)=\col(\Sigma_{xy}C)\subseteq\col(\Sigma_{xy}).
\label{eq:contain}
\end{equation}
Using $\Sigma_{yx}=\Sigma_{xy}^\top$,
\begin{equation}
F=\Sigma_{xy}\Sigma_{yy}^{-1}\Sigma_{xy}^\top\Sigma_{xx}^{-1}.
\label{eq:Fsym}
\end{equation}
Hence
\begin{equation}
\operatorname{rank}(F)
=\operatorname{rank}\!\bigl(\Sigma_{xy}\Sigma_{xy}^\top\bigr)
=\operatorname{rank}(\Sigma_{xy})
=: r,
\label{eq:rank}
\end{equation}
since $\operatorname{rank}(AA^\top)=\operatorname{rank}(A)$ for any $A$.
By \eqref{eq:contain} the subspace $\col(F)$ is contained in $\col(\Sigma_{xy})$,
and \eqref{eq:rank} shows they have the same dimension $r$. Therefore
$\col(F)=\col(\Sigma_{xy})$.
\end{proof}

Next, we order the reconstructed images along the visual feature of interest (eg. textured vs smooth images). This is done using heuristics described in the Appendix. EEG patterns for all images in each group are averaged, to form the corresponding group EEG pattern (eg. texture pattern vs smooth pattern). Using this procedure, we produce maps for the following visual features: visual semantics (animal vs food vs other), texture (textured vs smooth), hue (red vs blue) and brightness (bright vs dark). 
In the case of semantic EEG patterns, to minimize color-related confounds, we first grayscaled all images before extracting CLIP latents from them (see Fig. \ref{fig:recon_grayscale}). 

\subsubsection{EEG-fMRI pattern validation}
We repeat the analysis with the NSD dataset to create corresponding fMRI patterns which we compare spatially to the latent-filtered EEG patterns.

\paragraph{Representational Similarity Analysis (RSA)}
To compare the information encoded in different representations (EEG, CLIP, etc), we use representational similarity matrices (RSMs; \cite{kriegeskorteRepresentationalSimilarityAnalysis2008}) generated using the Pearson correlation coefficient between EEG and EEG Patterns corresponding to different stimulus features (e.g., semantic class, texture, luminance, etc.). 


\section{Results}

\subsection{Reconstruction performance}
\begin{figure}
    \centering
    \includegraphics[width=0.6\linewidth]{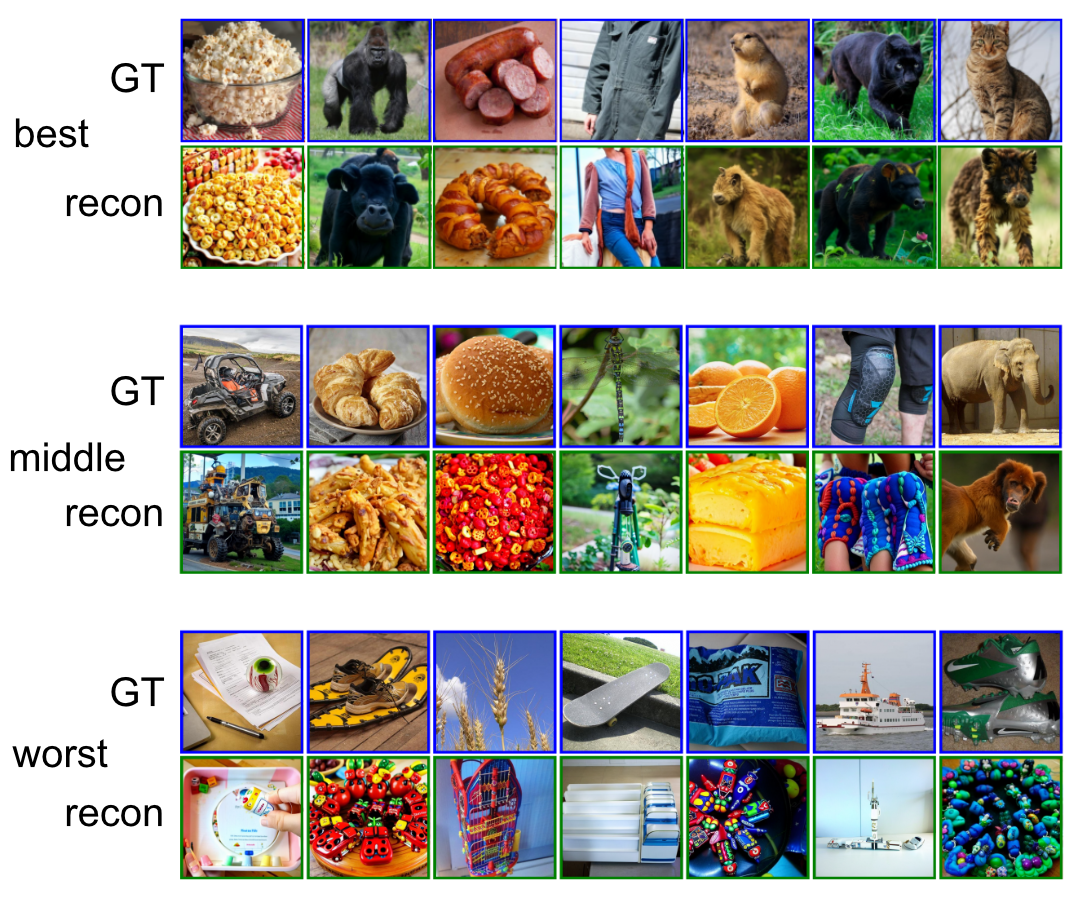}
    \caption{Reconstruction examples from Subject 1 using the CLIP latent space and Versatile Diffusion reconstructer, categorized into best, middle and worst. Best examples were selected by visual inspection, and middle and worst examples were selected by a CLIP score ranking of 94-100 and 194-200 respectively. The rows labeled GT and recon refer to ground truth and reconstructed images respectively. (For full reconstructions, see Fig. \ref{fig:recon_plot_ordered_by_performance}) }
    \label{fig:examples}
\end{figure}

\subsubsection{From CLIP}
The Versatile Diffusion Pipeline produces reconstructions consistent with the stimulus images in various aspects such as color, texture, and semantic meaning (see Fig. \ref{fig:examples}). The reconstruction performance of our simple linear model (shown quantitatively in Table \ref{tab:performance} achieves state-of-the-art reconstruction performance on all the standard metrics. Individual subject performance and full reconstruction examples are provided in the Appendix.

\begin{figure}[h] 
    \centering
    \includegraphics[width=0.7\linewidth]{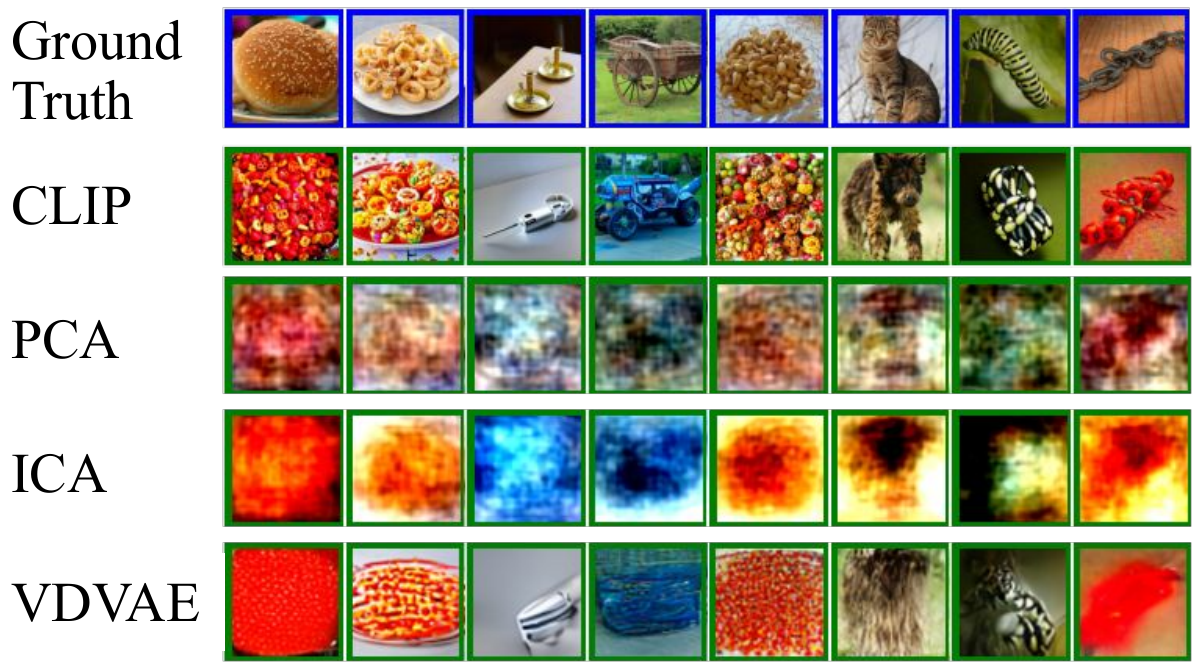}
    \caption{Ground truth stimulus images shown at the top; and reconstructions using different latent spaces, in the order: CLIP, PCA, ICA, and VDVAE.}
    \label{fig:low-level}
\end{figure}

\begin{table*}[h] 
\centering
\caption{Quantitative assessments of the reconstruction quality for EEG, MEG, and fMRI. For our algorithm we give the mean and standard deviation across 10 subjects with random seed 0. For detailed explanations of the metrics see section \ref{evaluation_metrics}.}
\label{tab:generation_comparison}
\resizebox{\textwidth}{!}{%
\begin{tabular}{lcccccccc}
\toprule
 & \multicolumn{4}{c}{Low-level} & \multicolumn{4}{c}{High-level} \\
\cmidrule(r){2-5} \cmidrule(l){6-9}
Dataset  & PixCorr $\uparrow$ & SSIM $\uparrow$ & AlexNet(2) $\uparrow$ & AlexNet(5) $\uparrow$ & Inception $\uparrow$ & CLIP $\uparrow$ & EffNet $\downarrow$ & SwAV $\downarrow$  \\
\midrule
NSD (Brain-Diffuser) \cite{Ozcelik_VanRullen_2023} & 0.254 & 0.356  & 0.942  & 0.962 & 0.872 & 0.915 & 0.775 & 0.423 \\
NSD (MindEye) \cite{Scotti_Banerjee_Goode_Shabalin_Nguyen_Cohen_Dempster_Verlinde_Yundler_Weisberg_et_al_2023} & $0.309$ & $0.323$  & $0.947$  & $0.978$ & $0.938$ & $0.941$ & $0.645$ & $0.367$ \\
\bf{NSD (Perceptogram with unCLIP, excluding sub-1)}  & $0.227 \pm  .008$ & ${0.339} \pm 0.003$ & ${ 0.894} \pm 0.013$ & ${0.946} \pm 0.011$ & ${0.883} \pm 0.0017$ & ${ 0.922} \pm 0.008$ & $0.759 \pm 0.018$ & ${ 0.405} \pm 0.009$ \\
\midrule
THINGS-MEG (BrainDecoding) \cite{Benchetrit_Banville_King} & $0.088$ & $0.333$ & $0.747$ & $0.855$ & $0.712$ & $0.804$ & - & $0.576$ \\
THINGS-MEG (EEGImageDecode) \cite{Li_Wei_Li_Zou_Liu_2024} & - & $0.340$  & $0.613$  & $0.672$ & $0.619$ & $0.603$ & - & $0.651$ \\
\bf{THINGS-MEG (Perceptogram with unCLIP)}  & $0.187 \pm  .004$ & ${\bf 0.376} \pm 0.007$ & ${\bf 0.848} \pm 0.036$ & ${\bf 0.906} \pm 0.031$ & ${\bf 0.748} \pm 0.032$ & ${\bf 0.826} \pm 0.027$ & $0.875 \pm 0.021$ & ${\bf 0.527} \pm 0.021$ \\
\midrule
THINGS-EEG2 (EEGImageDecode) \cite{Li_Wei_Li_Zou_Liu_2024}  & - & $0.345$ & $0.776$  & $0.866$ & $0.734$ & $0.786$ & - & $0.582$ \\
\bf{THINGS-EEG2 (Perceptogram with Versatile Diffusion)}& $0.267 \pm  .015$ & ${0.347} \pm 0.003$ & ${\bf 0.910} \pm 0.010$ & ${\bf0.927} \pm 0.005$ & ${\bf0.752} \pm 0.008$ & ${\bf 0.807} \pm 0.009$ & $0.877 \pm 0.004$ & ${ 0.540} \pm 0.004$ \\
\bf{THINGS-EEG2 (Perceptogram with unCLIP)}& $0.223 \pm  .029$ & ${\bf0.37} \pm 0.005$ & ${ 0.875} \pm 0.013$ & ${0.915} \pm 0.008$ & ${0.749} \pm 0.024$ & ${ 0.806} \pm 0.016$ & $0.87 \pm 0.011$ & ${\bf 0.530} \pm 0.009$ \\

\bottomrule
\end{tabular}%
}
\label{tab:performance}
\end{table*}

\subsubsection{Reconstruction from other latent spaces}
Observing the exemplar reconstructions shown in Fig. \ref{fig:low-level}, we observe the following trends for each latent space: Reconstructions from PCA primarily capture brightness of the original stimuli. The red versus blue hue is well captured by ICA, as the reds and blues are saturated in the reconstructions and are consistent with the warmth or coldness of the ground truth images. And as mentioned previously, reconstructions from VDVAE latents capture the level of texture in the stimuli. (\texttt{calamari}, \texttt{pistachios}, and \texttt{cheetah} look visually ``busy", while \texttt{CD player} and \texttt{cheese} look smooth). 

\subsubsection{Reconstruction after Electrode mirroring}
\begin{figure}
    \centering
    \includegraphics[width=0.75\linewidth]{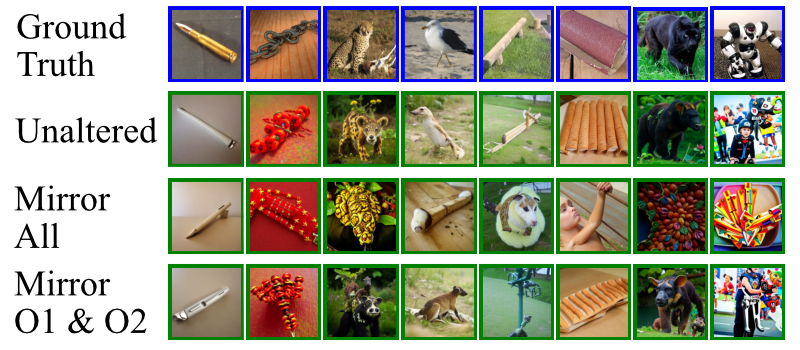}
    \caption{Examples of reconstructions for different test-time manipulations. \textbf{Ground Truth}: ground truth stimulus images; \textbf{Unaltered}: Unaltered Versatile Diffusion reconstructions; \textbf{Mirror All}: Pipeline trained normally, and electrode locations mirrored about the midline during test time (e.g. data from electrodes on the right scalp mapped to channels trained with data from electrodes on the left side); \textbf{Mirror  O1 \& O2}: Pipeline trained normally, but O1 and O2 are swapped during test time.} 
    \label{fig:mirroring}
\end{figure}

Reconstructions from unaltered and mirrored data are shown in Fig. \ref{fig:mirroring}.
(a) In the ``Mirror-all" condition, the reconstructions are altered both visually and semantically. Compared to the unaltered condition, we found that many images reversed their animacy. For example, \texttt{cheetah}, \texttt{seagull}, \texttt{panther}, and \texttt{robot} all produced non-living objects. Conversely, \texttt{balance beam} and \texttt{sandpaper} produced mirrored reconstructions that look like living creatures. (b) The ``mirror O1 \& O2" condition exhibited low-level visual changes, but largely preserved the semantic meaning obtained in the unaltered condition. In both mirror conditions, we noticed that some simple stimuli show a change in the angle or orientation of the reconstruction (eg. \texttt{bullet} and \texttt{chain}) consistent with a swapping of right and left visual field of the reconstructed image. Although the examples shown here are hand-picked to demonstrate these findings, full reconstructions for the mirroring conditions are shown in the appendix, and there are more examples there such as \texttt{coverall}, \texttt{pig}, \texttt{sausage}, \texttt{magician hat}, etc. 

\subsubsection{Time-Swapping} 
\begin{figure*}
    \centering
    \includegraphics[width=\linewidth]{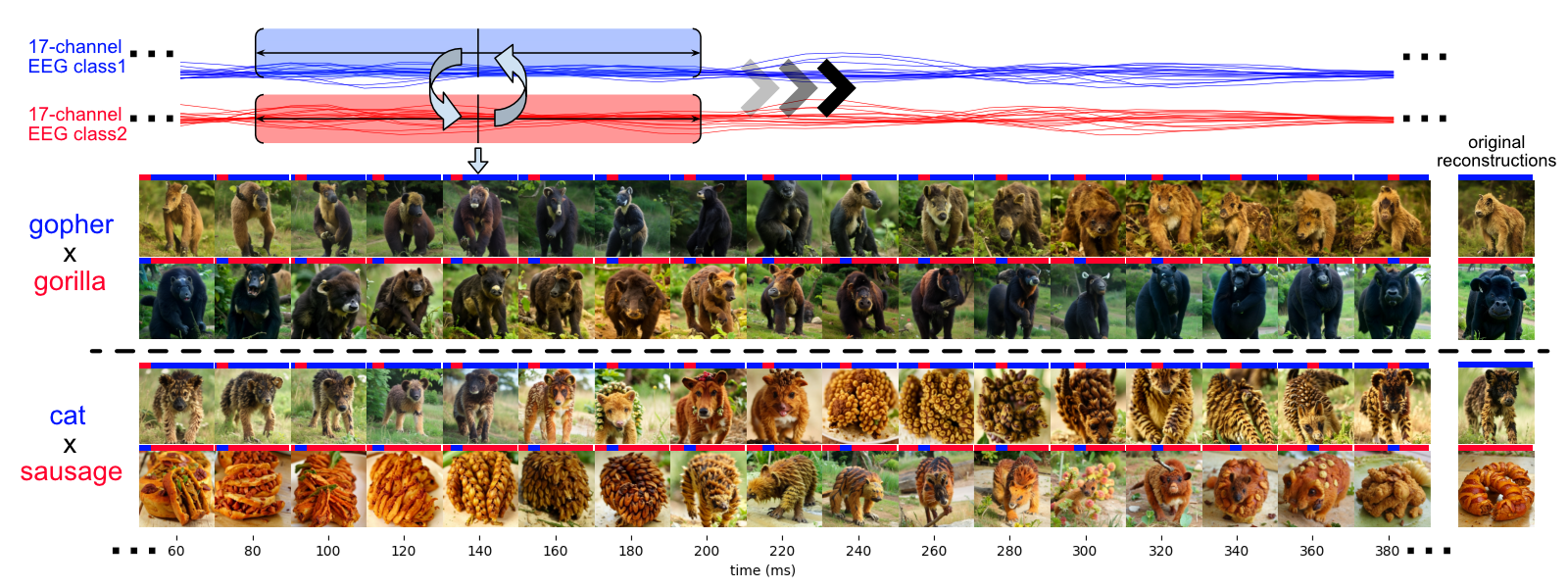}
    \caption{Illustration of the Time-Swapping Experiment.
    \textbf{Top}:  illustrates time segment swapping as a sliding window with the down arrow pointing to the corresponding reconstruction; \textbf{Bottom}: bar color over each image illustrates proportionally which segments come from its own EEG  and which comes from EEG to  the other class.}
    \label{fig:appendix_timeswap}
\end{figure*}

In the gopher-gorilla swap experiment shown in Fig. \ref{fig:appendix_timeswap}., the reconstructed “gopher” image has darker fur when the swapped windows are centered at 100ms through about 260ms (when 120ms time windows from 100-60=40ms to 260+60=320ms are replaced with the EEG to the gorilla from the corresponding time frame). Similarly the gorilla has a lighter fur color when the EEG in about the same time range is replaced with the EEG from the gopher presentation. In the cat-sausage swap experiment, the cat reconstruction has a food-like appearance when 120ms windows centered from 240-280ms and the sausage has an animal-like appearance when 120ms windows centered from 200-360ms are replaced with EEG from the cat presentation. The later sensitive time period for the semantic differences (animal vs. food) compared to the fur color differences (light vs. dark) reveals later processing of semantic compared to low-level visual features. 

\subsection{Decoding performance}
We trained a linear regression with no bias term from zero-mean EEG data to CLIP embeddings. We also trained a linear network with single layer from EEG dimension to CLIP dimension directly with no bias using MSE loss. The predicted CLIP embeddings were divided by their own L2-norm and multiplied by the mean training CLIP L2-norm. Both model were evaluated using the retrieval metric based on recall@1 and recall@5. 
\begin{table}[h]
    \centering
    \caption{Retrieval performance of EEG-to-CLIP decoding model.}
    \label{tab:retrieval_results}
    \begin{tabular}{@{}lcc@{}}
        \toprule
        Method & Recall@1 (\%) & Recall@5 (\%) \\
        \midrule
        Linear Regression & 25.0 & 50.0 \\
        Linear network & 24.5 & 49.5 \\
        \bottomrule
    \end{tabular}
\end{table}
\subsection{Encoding Weight Reconstructions Visualizes Presumptive Neural Preference}
\begin{figure}
    \centering
    \includegraphics[width=1\linewidth]{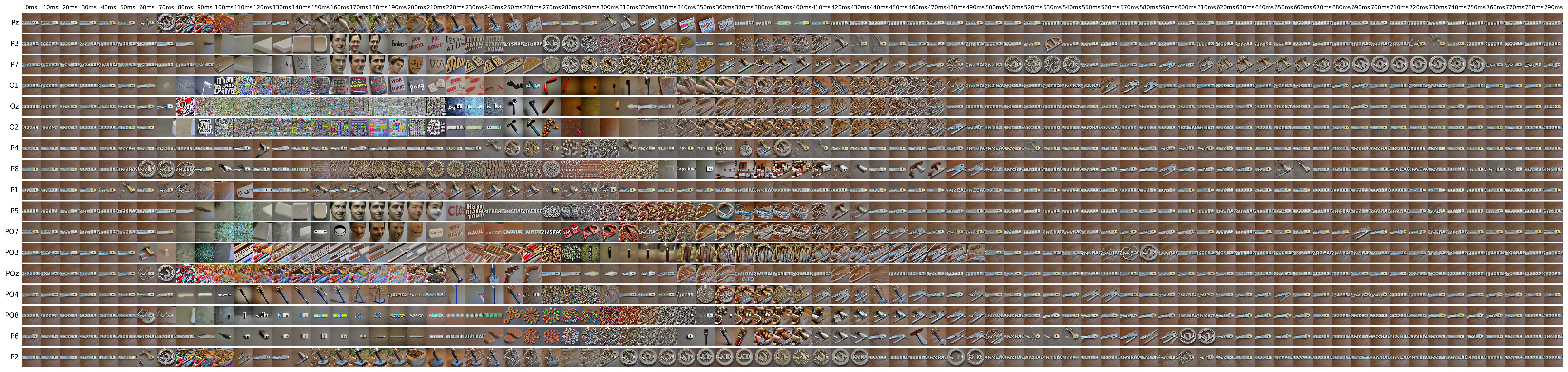}
    \caption{Encoding Weight Reconstructions Visualizes Presumptive Neural Preference (Subject 1). Each row represents an electrode in the visual cortical area and each column represents a 10ms time bin. Each individual image is a visual reconstruction of the linear encoding weight from CLIP to that particular electrode-time-bin, representing the CLIP visual features that would maximize the voltage at that point. But because here the linear weights are multiplied by -1, in this example it would mean the visual features that would minimize the voltage at that point. This would allow for the interpretation of negative ERP components. For example, there are 2 clusters of face images for electrode P3, P7, P5, and PO7 at around 160-200ms, so the face CLIP features would minimize the voltage for those electrodes around that time. The lateral location of these electrode and timing matches the N170 ERP component, which is a negative going ERP component around 170ms sensitive to faces. Because assigning meaning to either polarity involves somewhat of a subjective decision, we show in the next figure the positive weight reconstructions just as a comparison.}
    \label{fig:encoding_weights_full_grid_sub-01_inv_logmasked}
\end{figure}

\begin{figure}
    \centering
    \includegraphics[width=1\linewidth]{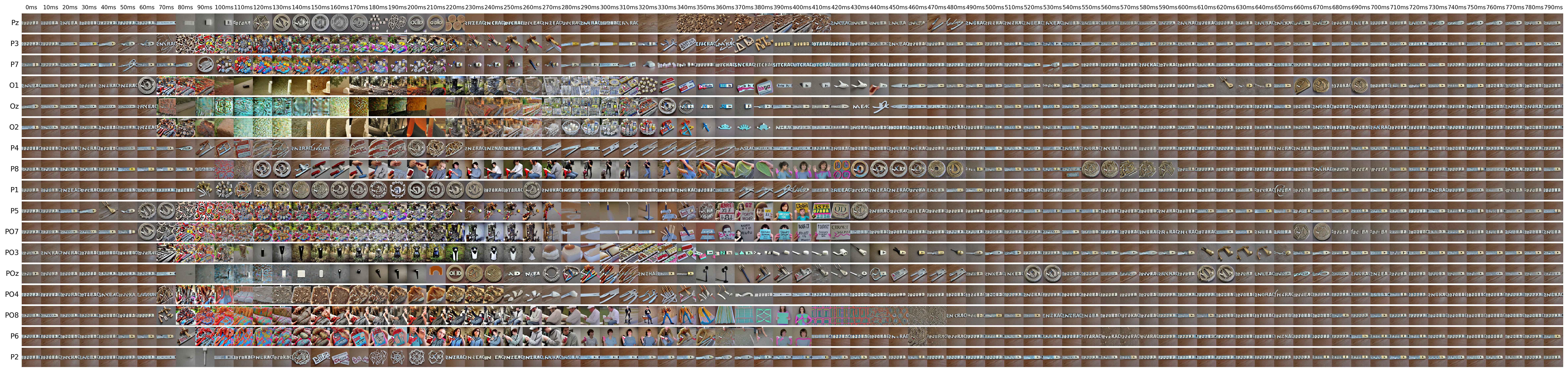}
    \caption{The positive weight version of Presumptive Neural Preference (Subject 1). Where the face cluster used to be for the negative weight reconstructions are now replaced by some kind of scenes here.}
    \label{fig:encoding_weights_full_grid_sub-01_logmasked}
\end{figure}

To directly visualize what visual features drive EEG responses, we reconstructed preferred stimuli from the CLIP-to-EEG encoding weights using unCLIP. This approach treats each weight vector as a CLIP embedding, revealing the "ideal stimulus" that would maximally activate each electrode according to our learned linear model.

The spatiotemporal reconstruction maps for subject 1 showed a high spatial frequency cluster for the medial channels such as O1, Oz, and O2 around 100-180ms, and a face cluster for the lateral channels such as P3, P7, P5, PO7 (in this case, all on the left-side) around 160-200ms. In this case, the weights are inverted (multiplied by -1), it is to show the CLIP features that give rise to negative voltages. The positive version of the weights are also included as a comparison.

\subsection{Spatiotemporal Pattern Analysis}

\subsubsection{Lateral vs Medial negativity encodes various visual features}
\begin{figure*}
    \centering
    \includegraphics[width=1\linewidth]{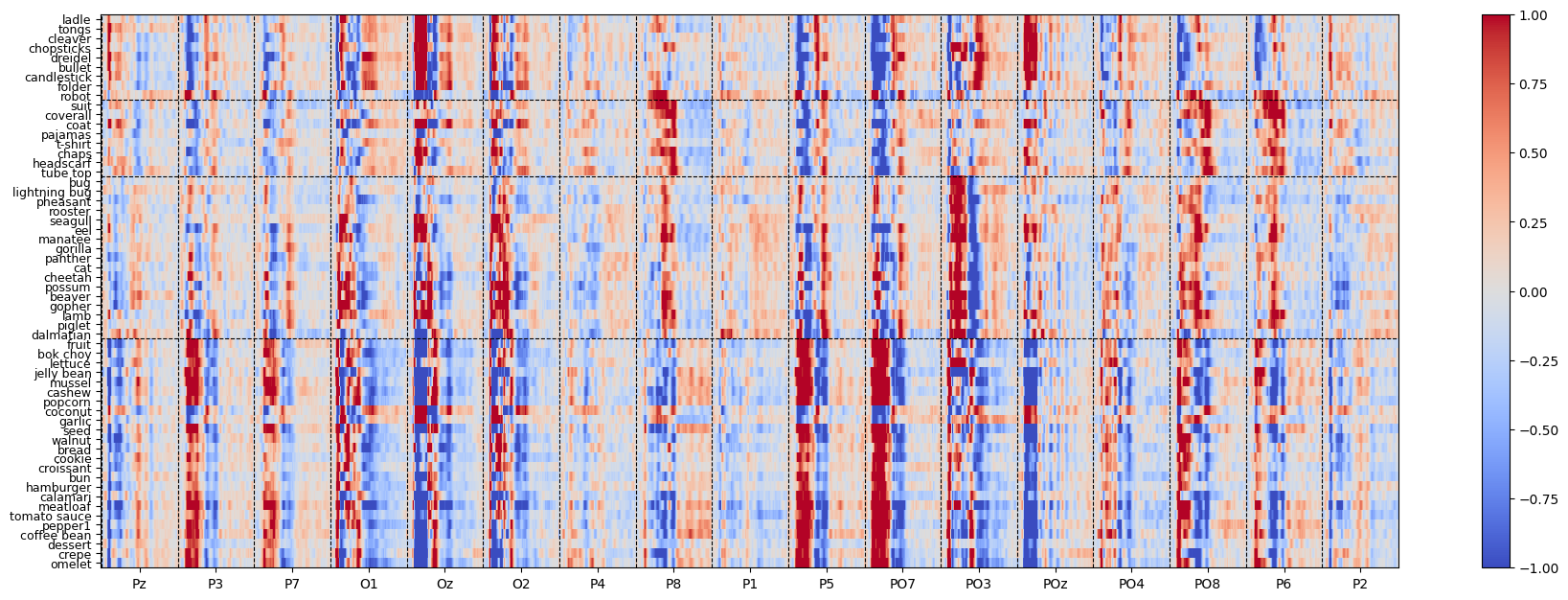}
    \caption{Selected EEG Patterns of CLIP (Subject 1). The hierarchical clustering on the CLIP embeddings extracted from the test images neatly organizes the 200 test categories into 3 general semantic groups (others, animals, food). Within the “others” group, it can be further subdivided into “small tools” and “clothing” with enough samples to see the pattern.}
    \label{fig:eeg_pattern_fewer2}
\end{figure*}

In this section, we contrast EEG patterns associated with various visual features. Accordingly, we discuss spatial differences in EEG patterns corresponding to semantic categories from CLIP, texture categories from VDVAE, hue categories from ICA, and brightness categories from PCA. 

\begin{figure}
    \centering
    \includegraphics[width=0.5\linewidth]{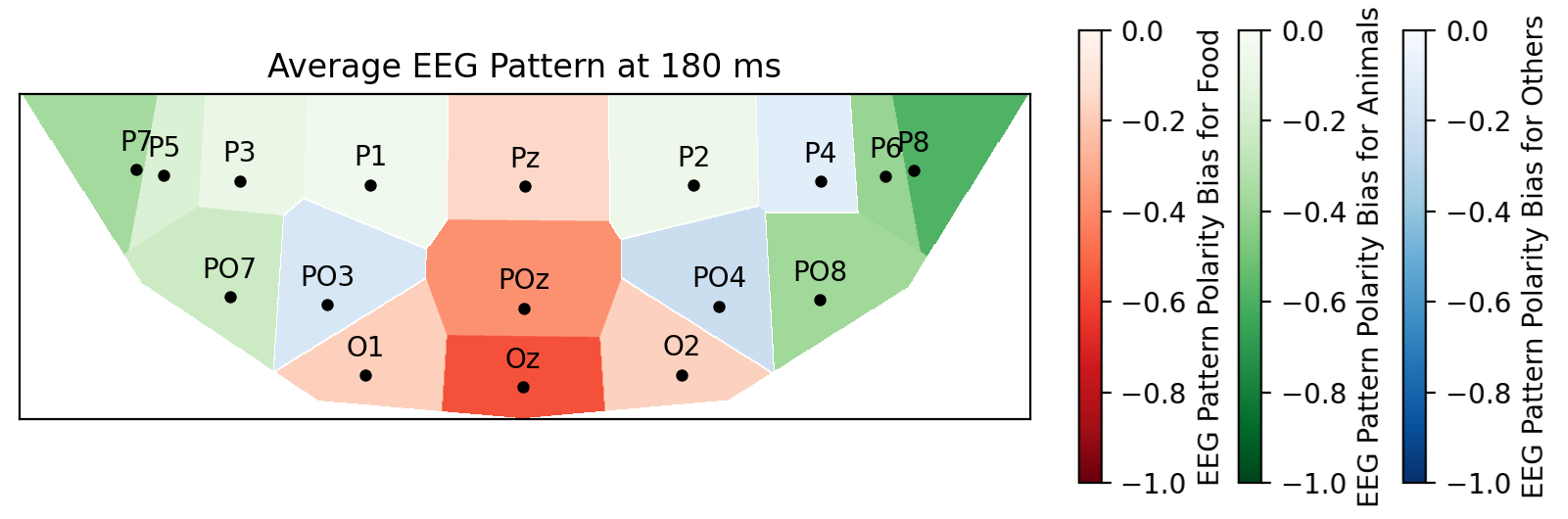}
    \caption{A temporal slice of 10-subject average spatiotemporal semantic map at 180ms for the ``food", ``animals" and ``others" categories. The figure shows electrode locations at the back of the head using the standard 10/10 naming system.
    Darker colors represent decreased voltage relative to the grand average response.
    }
    \label{fig:topography}
\end{figure}

\begin{figure}[h]
    \centering
    \includegraphics[width=0.9\linewidth]{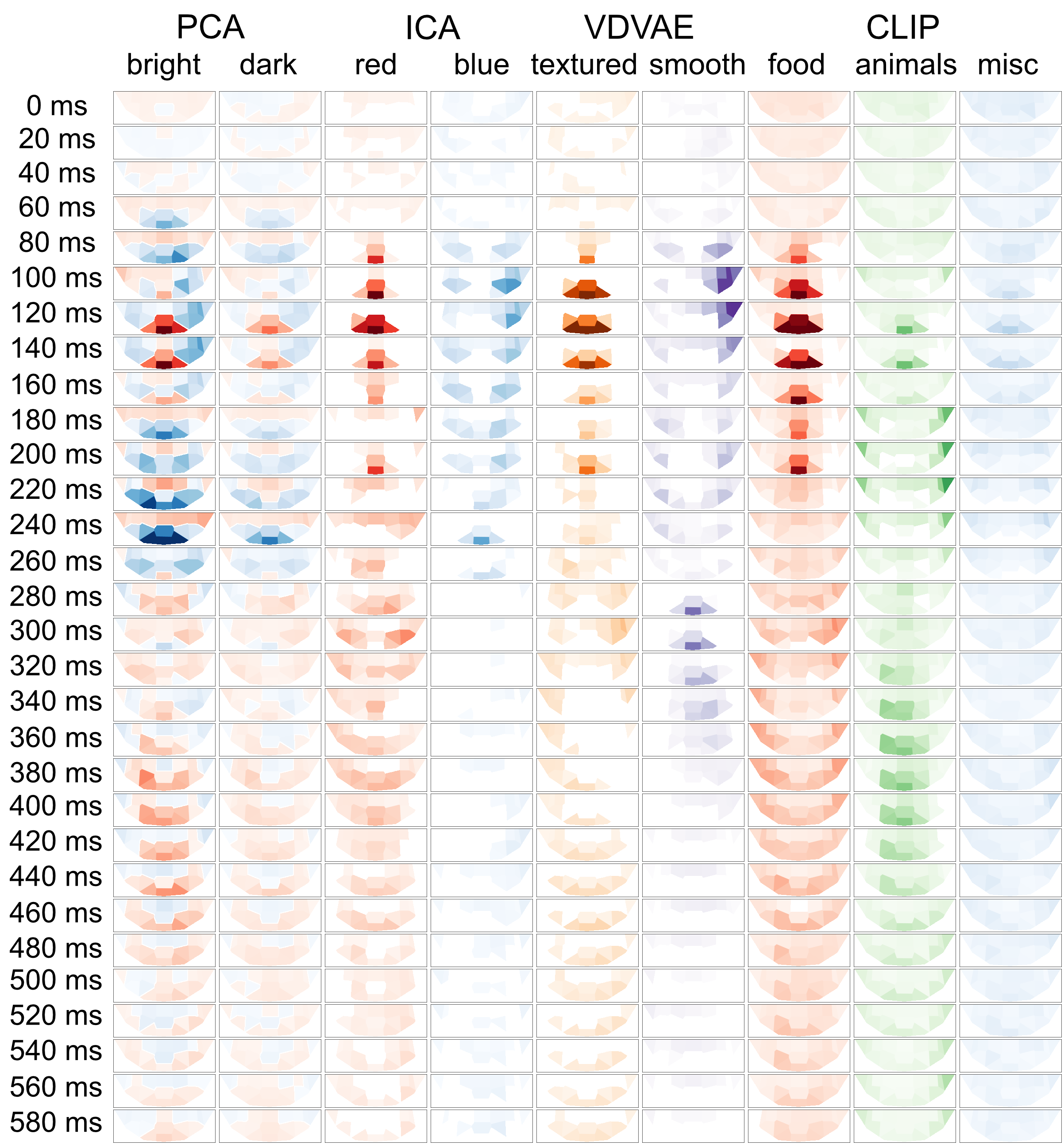}
    \caption{The 10-subject average EEG patterns of 4 different latent spaces: PCA, ICA, VDVAE, and CLIP. Each latent space is sub-divided into individual visual features that it preserves. For PCA, the red color means stronger positive polarity and blue means stronger negative polarity. For the other 3 latents, the stronger the color means the stronger the negative polarity. The negative polarity is chosen because the EEG has a negative-going peak around 100-200ms, and thus more negative around this time implies a stronger signal.
    } 
    \label{fig:eeg_pattern_topo_sub-avg}
\end{figure}

Fig. \ref{fig:topography} shows one temporal slice displaying a spatial contrast between three semantic categories (animals, food, other), plotted on a 2-D topological scalp map. While these maps are inspired by the fMRI semantic maps from \cite{Huth2016}, the temporal sensitivity of EEG permits the visualization of spatiotemporal maps that unfold in time, shown in Fig. \ref{fig:eeg_pattern_topo_sub-avg}. The semantic maps for individual subjects are shown in Fig. \ref{fig:indiv_eeg_patterns}. Most subjects show the same spatial pattern seen in the grand-average from Fig. \ref{fig:eeg_pattern_topo_sub-avg}, with some individual differences in asymmetry and lateralization strength. The individual assymetries likely underlie the performance degradation we observe earlier from mirroring the electrodes.

The EEG patterns from VDVAE and ICA show similar medial vs. lateral spatial separation as the patterns from CLIP (see Fig. \ref{fig:eeg_pattern_topo_sub-avg}). Concretely, EEG patterns from VDVAE show that smoother reconstructions have more lateral negativity, and more textured reconstructions are associated with more medial negativity. Similarly, EEG patterns from ICA indicate that cooler reconstructions are the result of more lateral negativity, while warmer images have more medial negativity. 

Finally, EEG patterns from PCA show that brighter reconstructions are the result of more medial positivity around 120ms, and more medial negativity around 240ms; while darker reconstructions are close to the grand average.

Note that all EEG patterns we obtain (except brightness-related ones) are negative-going from 100-200ms, so a larger negative value around this time implies a stronger signal. We thus show the negative-going half of the patterns, with a single color in each column (except in the case of EEG patterns from PCA, with red and blue indicating positive and negative polarities respectively). This implication about signal strength is important in the following section when we compare EEG patterns with fMRI patterns.

\subsubsection{Temporal Difference Between Low-level and High-level Patterns}
Note that the spatiotemporal maps for ``blue" and ``smooth" overlap spatially with ``animals" (see Fig. \ref{fig:eeg_pattern_topo_sub-avg}, where all three showed lateral negativity). The difference lies entirely in the timing: the lateral negativity for ``animals" starts at around 180ms, which is later than ``blue" and ``smooth" (which starts at around 100ms). This temporal difference would not be apparent in a static spatial map from fMRI, and is discussed further in the following sections.

\subsubsection{fMRI patterns show similar Lateral vs Medial differences}
\begin{figure*}
    \centering
    \includegraphics[width=1\linewidth]{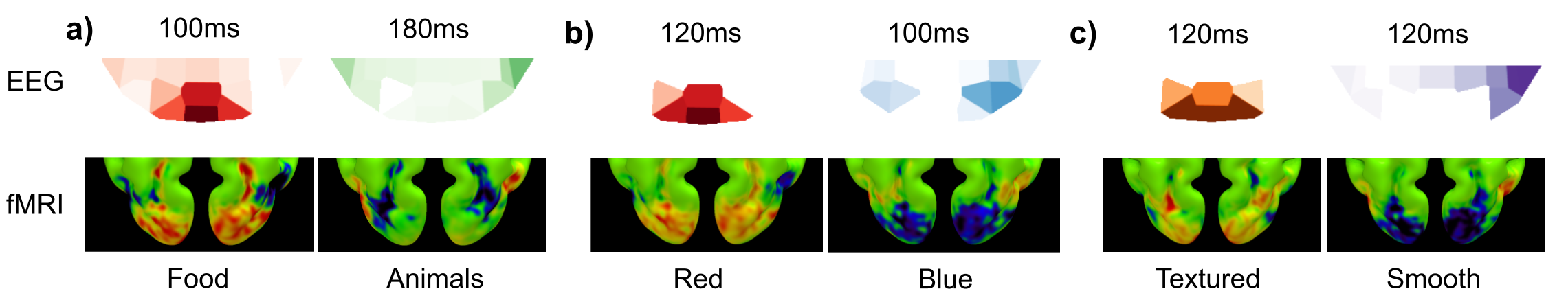}
    \caption{Cross-subject averaged EEG and fMRI patterns for different visual features. For the EEG maps, the colors of each pairing to the visual feature category. (a) food (red) and animals (green), (b) red (red) and blue (blue), and (c) textured (orange) and smooth (purple).  The fMRI maps use a standard fMRI color scheme (independent of category) where red represents increased BOLD signal and blue decreased. The increased fMRI signal (red) corresponds to increased EEG signal represented by the darker category-specific color.}
    \label{fig:three_patterns_abc}
\end{figure*}

Here we show that the primary difference of animate vs. inanimate is widely distributed as a ``lateral vs medial" spatial difference (see Fig. \ref{fig:three_patterns_abc}). This is the same spatial pattern in the negative voltage of EEG patterns.

In the context of fMRI, the patterns correspond to the relative brain activations of each category compared to the mean pattern. The fMRI patterns consistently separate in the medial versus lateral axis in much the same way as EEG for color and texture as well.

\begin{figure}[h]
    \centering
    \includegraphics[width=0.5\linewidth]{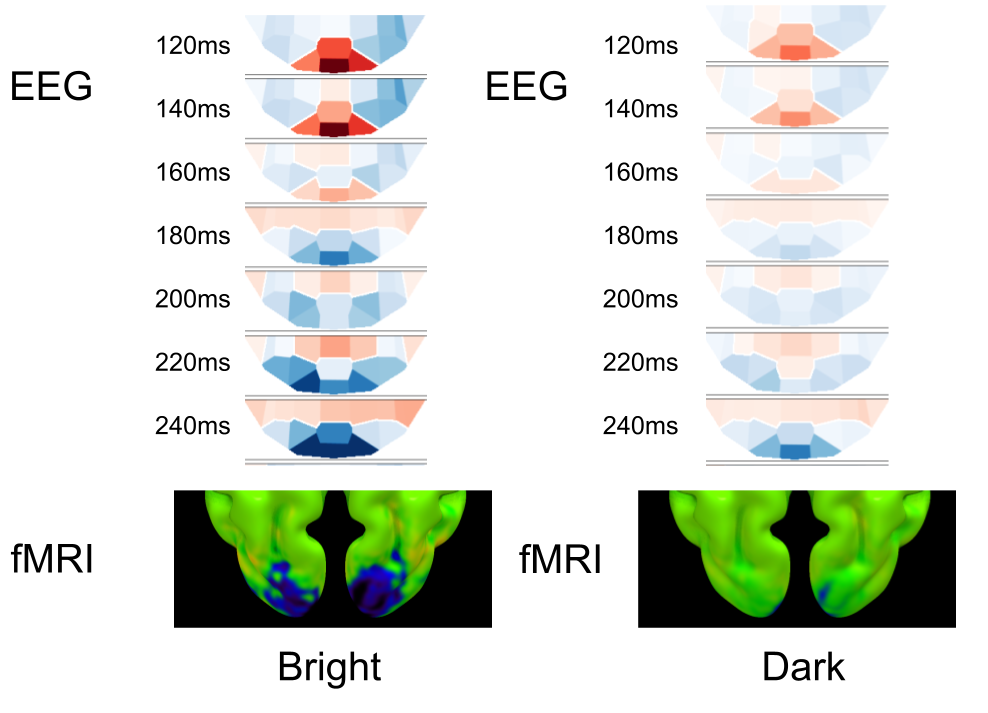}
    \caption{Cross-subject averaged EEG and fMRI patterns for ``bright" and ``dark". The EEG maps use the same color scheme as Fig. \ref{fig:eeg_pattern_topo_sub-avg} where red shows more positive voltage and blue more negative.  The fMRI maps use a standard fMRI color scheme (independent of category) where red represents increased BOLD signal and blue decreased. While fMRI shows decreased BOLD response to brighter stimuli, the EEG shows a stronger positive then negative pattern for brighter stimuli.}
    \label{fig:bright-vs-dark}
\end{figure}

Lastly, the fMRI pattern for brightness appears to be lower in the medial area for bright images, and roughly equal to the grand average for the dark images. Similarly, for the EEG, the patterns for both bright images and dark images occupy similar spatial locations with dark image pattern being much closer to the grand average (See Fig. \ref{fig:bright-vs-dark}). 

\section{Discussion}

We present a unified framework that not only enables image reconstruction from EEG using a simple linear decoder, but also isolates the latent-filtered EEG patterns that support visual perception. 
Our findings challenge the prevailing assumption that high-fidelity visual decoding from EEG requires deep, non-linear models. Instead, we demonstrate that EEG representations are inherently compatible with CLIP’s latent structure, enabling direct linear mapping for high-quality image reconstructions. 
Alignment with spatial patterns derived from fMRI further validates the spatial organization captured by our EEG-derived features, 
 while also highlighting the unique temporal dynamics accessible through EEG.
In the following two sections, we discuss (a) Why a simple linear mapping is sufficient to achieve the reported reconstruction performance, and (b) what our perturbation experiments and latent-filtered EEG patterns reveal about the spatial and temporal organization of visual representations in the EEG.

\subsection{Why does EEG visual reconstruction work?}
\begin{figure}[!hb]
    \centering
    \includegraphics[width=0.6\columnwidth]{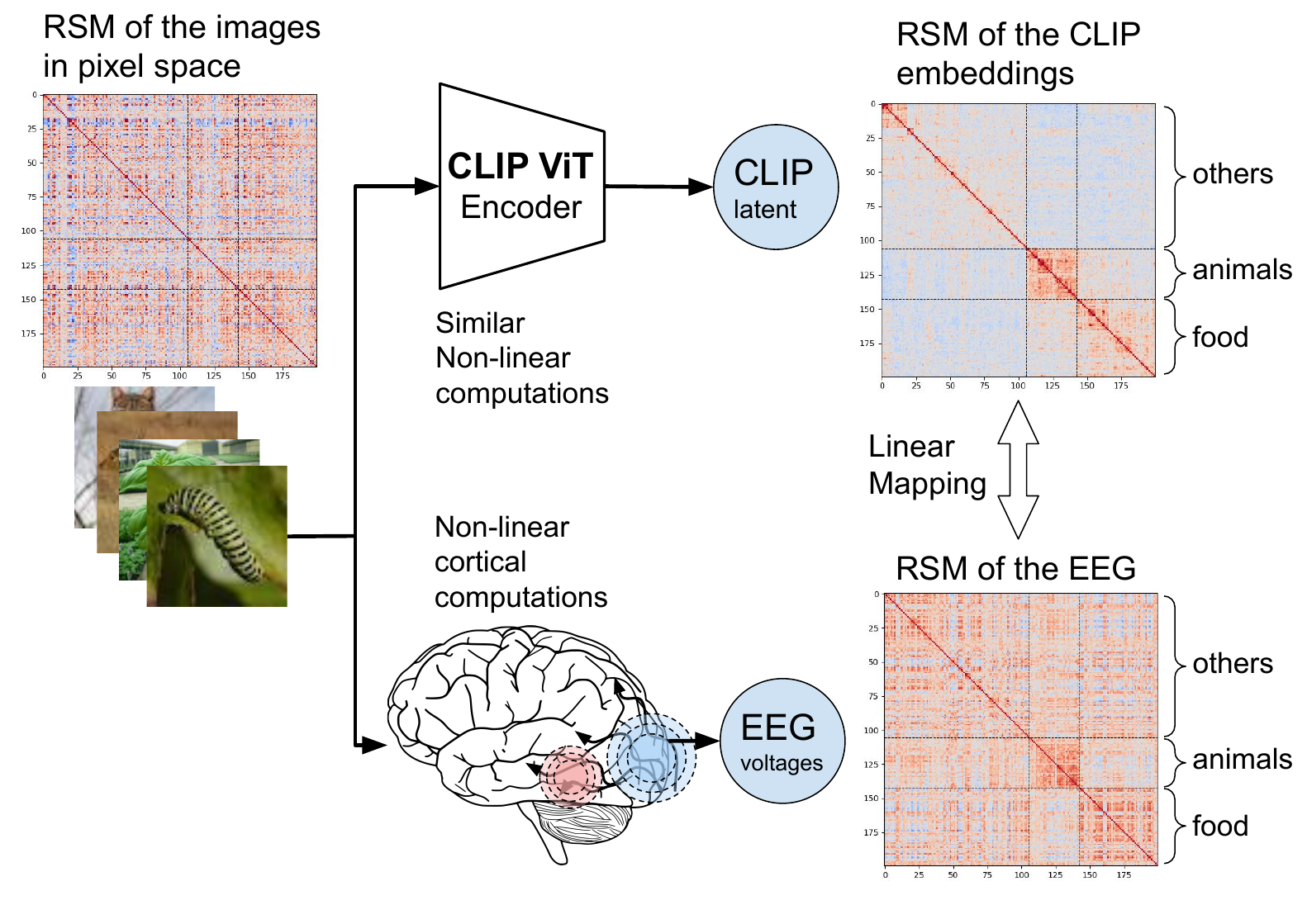}
    \caption{CLIP-image embeddings are generated by a large visual transformer (ViT-L/14), while the EEG activity from visual perception is generated by neural processing of visual stimuli summed with other internal and noise processes. EEG recording by nature of its summation of large-scale neural activity, is able to capture information from along the visual pathway including low-level pixel-related features as well as high-level semantic related features; these are all present in the EEG signal. Likewise the CLIP representation consists of representations of units from the CLIP hierarchical neural network. The RSMs show that the EEG and CLIP exhibit a shared representational structure with respect to the semantic class of the image.}
    \label{fig:s1}
\end{figure}

\begin{figure}
    \centering
    \includegraphics[width=0.5\linewidth]{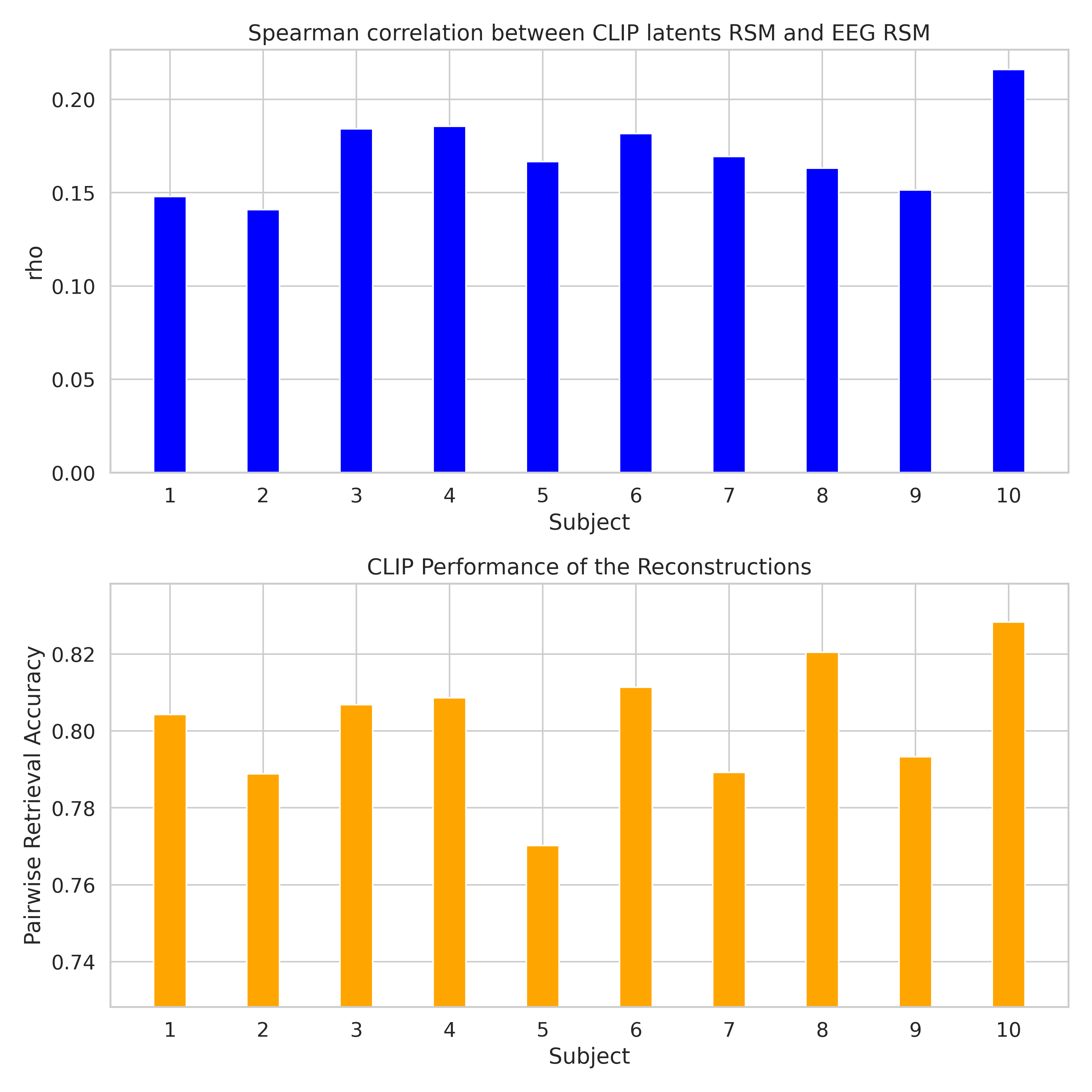}
    \caption{Subject-wise CLIP–EEG RSM similarity vs.\ reconstruction performance (Spearman $r = 0.579$).}
    \label{fig:spearman_similarity}
\end{figure}

Unlike deep learning methods that rely on multi-layered feature extraction, Perceptogram linearly maps EEG responses to CLIP latents.



To understand why EEG visual reconstruction is possible using a simple linear decoder, we examine the representational similarity matrices (RSMs) across EEG, CLIP, and pixel space (Fig.~\ref{fig:s1}). While pixel-based representations show little structure, both EEG and CLIP embeddings organize stimuli into clear semantic clusters (e.g., animals vs. food). This shared structure (average Spearman correlation between EEG and CLIP test RSMs is .22) suggests that both the brain and CLIP perform nonlinear transformations from raw visual input to a semantically structured latent space, enabling their linear alignment 
(see Fig.~\ref{fig:spearman_similarity}).  
Alignment has also been shown in previous comparisons between neural and deep network activations~\cite{Allen_St-Yves_Wu_Breedlove_Prince_Dowdle_Nau_Caron_Pestilli_Charest_etal._2022, tu2018relating, yamins2013hierarchical}.

\subsection{How does EEG visual reconstruction work?}

Our perturbation experiments and latent-filtered EEG patterns reveal key insights into how visual representations are encoded in EEG signals, both spatially and temporally.
In line with prior work, our time-swapping experiments suggest that visual information is distributed across distinct temporal windows. Specifically, low-level visual features such as fur color are encoded earlier, while higher-level semantic content—such as animacy—is encoded later in the EEG time course. 
Spatially, swapping only the O1 and O2 electrodes can flip reconstructions across the vertical midline, consistent with the known contralateral organization of early visual processing. In contrast, mirroring all electrodes induces broader changes, affecting both visual and semantic attributes in the reconstruction. Notably, some reconstructions originally depicting animate objects appear inanimate after mirroring, and vice versa. 

This finding, together with the subject-specific CLIP-filtered EEG spatial patterns associated with animals, provides converging evidence that animacy may be encoded asymmetrically and laterally at the individual level. In contrast, EEG responses to food appear more medially organized and less sensitive to mirroring. This medial-lateral distinction also corresponds to other visual features (e.g., texture, brightness), with 
individual differences in lateralization, as illustrated in Fig.~\ref{fig:indiv_eeg_patterns}.

\subsubsection{Animacy (Animals vs. Food)}
The semantic maps obtained in Fig. \ref{fig:eeg_pattern_topo_sub-avg} open up a potentially new way to interpret the voltage 100-200 ms after stimulus onset. A negative polarity bias for faces at the occipito-temporal electrodes between 100-200ms is commonly known as the N170. Past studies have indicated its sensitivity to non-face categories, and particularly categories that are highly familiar \citep{Rossion2008}. Here, we provide another potential explanation to the N170 component amplitude -- animacy. Animacy, being a common organizing principle in visual processing, involves activation in the ventral temporal cortex \citep{kriegeskorteRepresentationalSimilarityAnalysis2008}. 

\subsubsection{Relating EEG to fMRI}
While in EEG, it is generally controversial to claim that observed electrode voltages correspond to neural activities directly underneath it, the locations are more easily interpretable in fMRI. The consistency between EEG and fMRI spatial locations allows a better understanding of the locations we observe on the EEG patterns. We see that, indeed, the reason we are seeing a lateral negativity for ``animals" and medial negativity for ``food" is because we see higher BOLD activities in the corresponding areas.

In the fMRI dataset, categories such as ``faces" and ``humans from a distance" activate similar ventral lateral cortical area (see Fig. \ref{fig:human-closeup}, \ref{fig:human-distant}) as ``animals"; categories such as ``room interiors", ``urban scenes" (see Fig. \ref{fig:interiors}, \ref{fig:urban})  and ``food" do not. This presents the hypothesis that N170 is the EEG equivalent of the animacy axis. 

\subsection{Limitation: Texture and Color Covariation}
Our results show that the medial-lateral separation is also present beyond the animacy axis. Images that look textured show more negative voltage at the medial-occipital electrodes and more BOLD activation in that area; and likewise for images that have reddish hue. 

We should caution that the natural datasets we, and many others, use contain correlations in their visual statistics. For instance, objects such as food tend to be highly textured and warm-colored. Animals typically have green or blue-ish backgrounds, and are typically not as textured as food (e.g. strawberries and burger buns with sesame seeds on them). Thus there is a possibility that if the brain maps any of these correlated features, it may appear to similarly map the confounded features. (Note this is a problem for all reconstruction work with these datasets).

\subsection{Applications to computer vision} 
We have shown that we can successfully linearly project EEG onto the CLIP latent space with reasonable (and SOTA) reconstruction performance. At the same time, there are systematic discrepancies in the similarity structure of the  EEG and the CLIP image representations.  It is possible that the EEG contains meaningful (to humans) information not adequately captured in the CLIP representation and that the RSM of EEG patterns may be helpful as a teaching signal to train computer vision models to be more similar to human vision.

\newpage
\bibliographystyle{apalike}    
\bibliography{references}  

\newpage
\appendix
\section{Extended Methods}
\subsection{Dataset Details}
\label{appendix:dataset}
\cite{Gifford_Dwivedi_Roig_Cichy_2022} for EEG analysis, and the publicly available Natural Scenes Dataset (NSD) from \cite{allen2022massive} for our fMRI analysis to validate findings from the EEG analysis.

\subsubsection{THINGS-EEG2}
EEG data was collected while an image is presented for 100ms followed by a blank screen for 100ms before the next image. The image presentation order is pseudo-randomized across the entire image set. 10 subjects viewed the same 16740 images, of which the same 200 images are test images. We used the preprocessed version (\url{https://osf.io/anp5v/}), which has 17 posterior EEG channels compared to the 63 total channels in the raw dataset. The EEG was initially sampled at 1000Hz and down-sampled to 100Hz during the preprocessing. The only major filtering method applied during the preprocessing is Multi-Variate Noise Normalization (MVNN), which computes the covariance matrices of the EEG data (calculated for each time-point), and then averages them across image conditions and data partitions. The inverse of the resulting averaged covariance matrix is used to whiten the EEG data (independently for each session) \citep{Gifford_Dwivedi_Roig_Cichy_2022}. 
Trials are extracted from $-0.2$s to 0.8s relative to the onset of the stimulus. Each training image is shown 4 times, and each test image is shown 80 times. We averaged all trials for the same image (within subject) to form the final dataset. At 100Hz sampling rate, $-0.2$s to 0.8 seconds corresponds to 100 samples. We discarded the first 20 samples which correspond to $-0.2$s to 0s, leaving 80 samples times 17 channels or 1360 dimensions per image per subject.  The final dimensions of the training data for each subject are (16540 images, 1360 features) for the training set, and (200 images, 1360 features) for the test set.

\subsubsection{NSD}
The fMRI data was collected while an image is presented for 3 seconds followed by a blank screen for 1 seconds before the next images. The images are taken from Microsoft's COCO image database \cite{coco_dataset_2014} rather than the THINGS initiative image database. The image presentation order is pseudo-randomized across the entire image set: 982 images (used as the test set) are shared across 4 trial-complete subjects (1, 2, 5, 7) while each subject has 8859 images (used as the training set)  exclusive to the particular subject and not shared across subjects. Each image is presented 3 times to enhance the signal-to-noise ratio, and the presentation order of which is controlled. 

\subsection{Model Architecture Details}
\begin{figure*}
    \centering
    \includegraphics[width=1\linewidth]{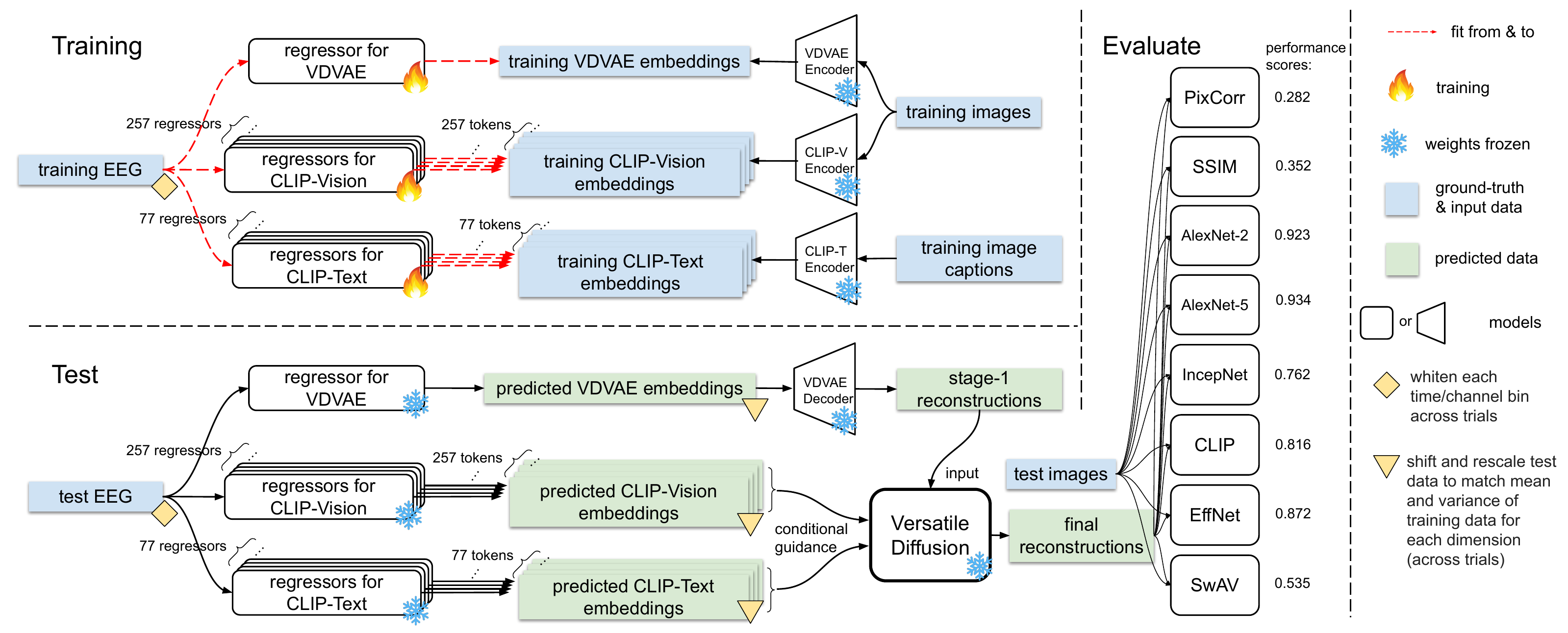}
    \caption{Flowchart illustrating the Versatile Diffusion variant of our reconstruction pipeline. In the \textbf{training} stage, images are processed through a VDVAE encoder, generating VDVAE embeddings (91168 dimensions), and through a CLIP-Vision encoder, producing CLIP-Vision embeddings (257 tokens × 768 dimensions). Corresponding captions are encoded by a CLIP-Text encoder to form CLIP-Text embeddings (77 tokens × 768 dimensions). Because CLIP-Vision and CLIP-Text embeddings have multiple tokens, separate regressors are trained to project the EEG data to each of the corresponding token space. Dashed red arrows indicate the fitting of these regressors from EEG to the embeddings, and only one arrow is shown for clarity since all regressors use the same EEG data.}
    \phantomsection
    \label{fig:flowchart_VD}
\end{figure*}

\begin{figure*}
    \centering
    \includegraphics[width=1\linewidth]{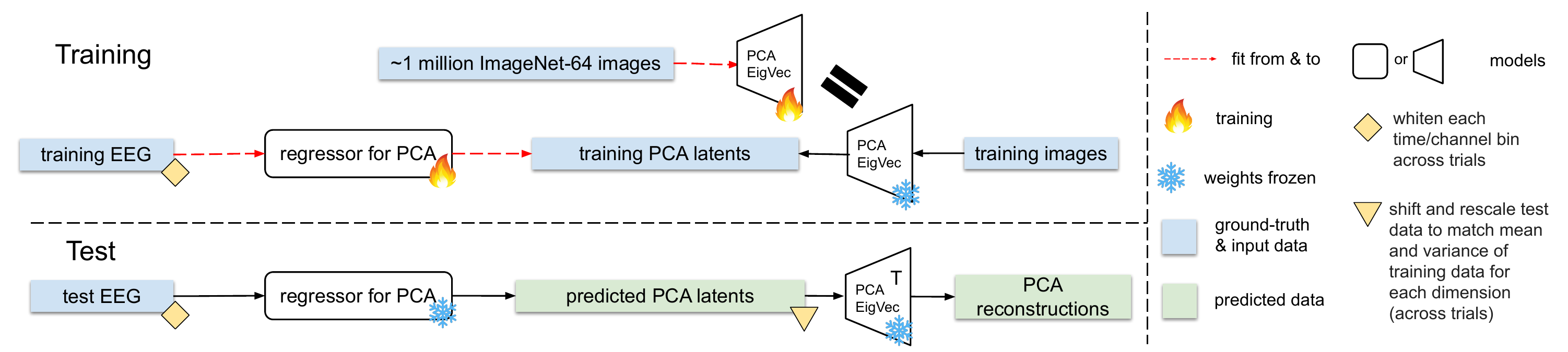}
    \caption{Flowchart illustrating the PCA reconstruction pipeline.}
    \label{fig:enter-label}
\end{figure*}

\begin{figure*}
    \centering
    \includegraphics[width=1\linewidth]{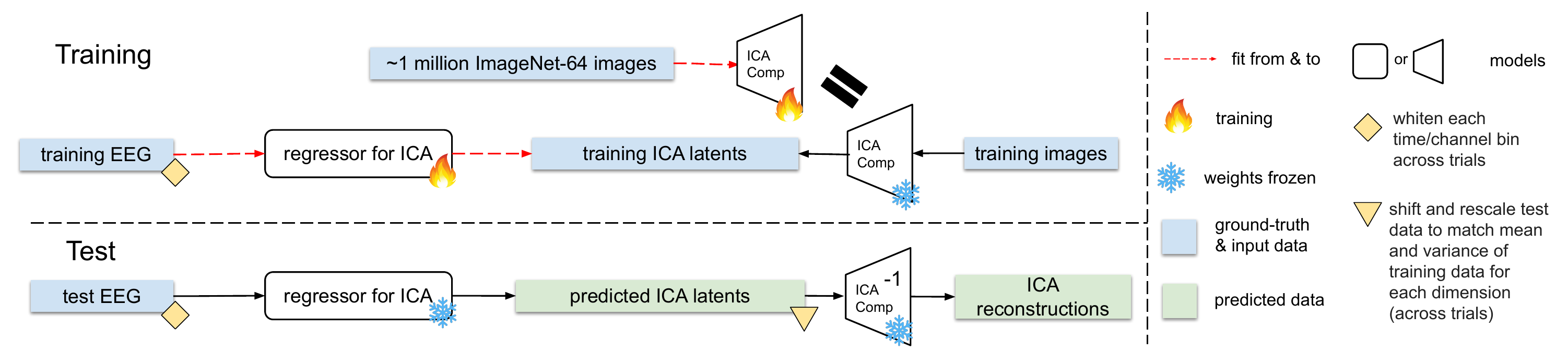}
    \caption{Flowchart illustrating the ICA reconstruction pipeline.}
    \label{fig:enter-label}
\end{figure*}

\subsection{Evaluation Metrics}
\label{evaluation_metrics}
We used the same performance metrics 
(see Table \ref{tab:performance}) as in \cite{Ozcelik_VanRullen_2023}, which has been used in other followup studies such as MindEye \cite{Scotti_Banerjee_Goode_Shabalin_Nguyen_Cohen_Dempster_Verlinde_Yundler_Weisberg_et_al_2023}. The 8 metrics we used are Pixel Correlation (PixCorr), Structural Similarity (SSIM), AlexNet layer 2 and 5 outputs pairwise correlations, InceptionNet output pairwise correlation, CLIP ViT output pairwise correlation, EfficientNet output distance, and SwAV output distance.
PixCorr and SSIM involve comparing the reconstructed image with the ground-truth (GT) test image. PixCorr is a low-level (pixel) measure that involves vectorizing the reconstructed and GT images and computing the correlation coefficient between the resulting vectors. SSIM is a measure developed by Wang et al. 2004 that computes a match value between GT and reconstructed images as a function of overall luminance match, overall contrast match, and a ``structural'' match which is defined by normalizing each image by its mean and standard deviation. We should note that this measure was designed for comparing images with minor distortions and does not seem as reliable for images that are not close matches as currently obtained with image reconstruction methods. One way to see this is to observe 
that the SSIM measure does not seem much affected by the duration of EEG window over 50ms, unlike the other measures that show performance improvements from 100ms through 400ms.

\subsection{Analysis Methods Details}

\subsubsection{Time-Swap Effect}
In order to investigate the temporal dynamics and the salient features in the EEG data, we develop a novel technique to find time-ranges that are most sensitive to disturbance.  We used pairs of images and swapped analogous time segments of data between EEG responses to each of the images.

Each image results from reconstruction of EEG where a 120ms time window centered at the corresponding time point is swapped between the 2 classes within that window while holding the signal outside the window the same. On top of each reconstructed image, we added a color bar that proportionally indicates which EEG time segment is swapped with the other class for that image. The two classes are represented by red and blue in this color bar and time is represented in the horizontal direction so a blue bar with a small red square represents that 120ms of the EEG at its relative location is swapped with the EEG for the other class. Notice how the small squares progress to the right as the samples progress to the right. The original, unswapped reconstructions (shown at right) have their color bars all blue/red, indicating that no part of their EEG is swapped with the other class.

In the gopher-gorilla swap experiment, the reconstructed “gopher” image has darker fur when the swapped windows are centered at 100ms through about 260ms (when 120ms time windows from 100-60=40ms to 260+60=320ms are replaced with the EEG to the gorilla from the corresponding time frame). Similarly the gorilla has a lighter fur color when the EEG in about the same time range is replaced with the EEG from the gopher presentation. In the cat-sausage swap experiment, the cat reconstruction has a food-like appearance when 120ms windows centered from 240-280ms and the sausage has an animal-like appearance when 120ms windows centered from 200-360ms are replaced with EEG from the cat presentation. The later sensitive time period for the semantic differences (animal vs. food) compared to the fur color differences (light vs. dark) reveals later processing of semantic compared to low-level visual features.

\subsection{Heuristics for ordering images along a visual feature}
\label{appendix:ordering_heuristics}
\begin{figure}
    \centering
    \includegraphics[width=1\linewidth]{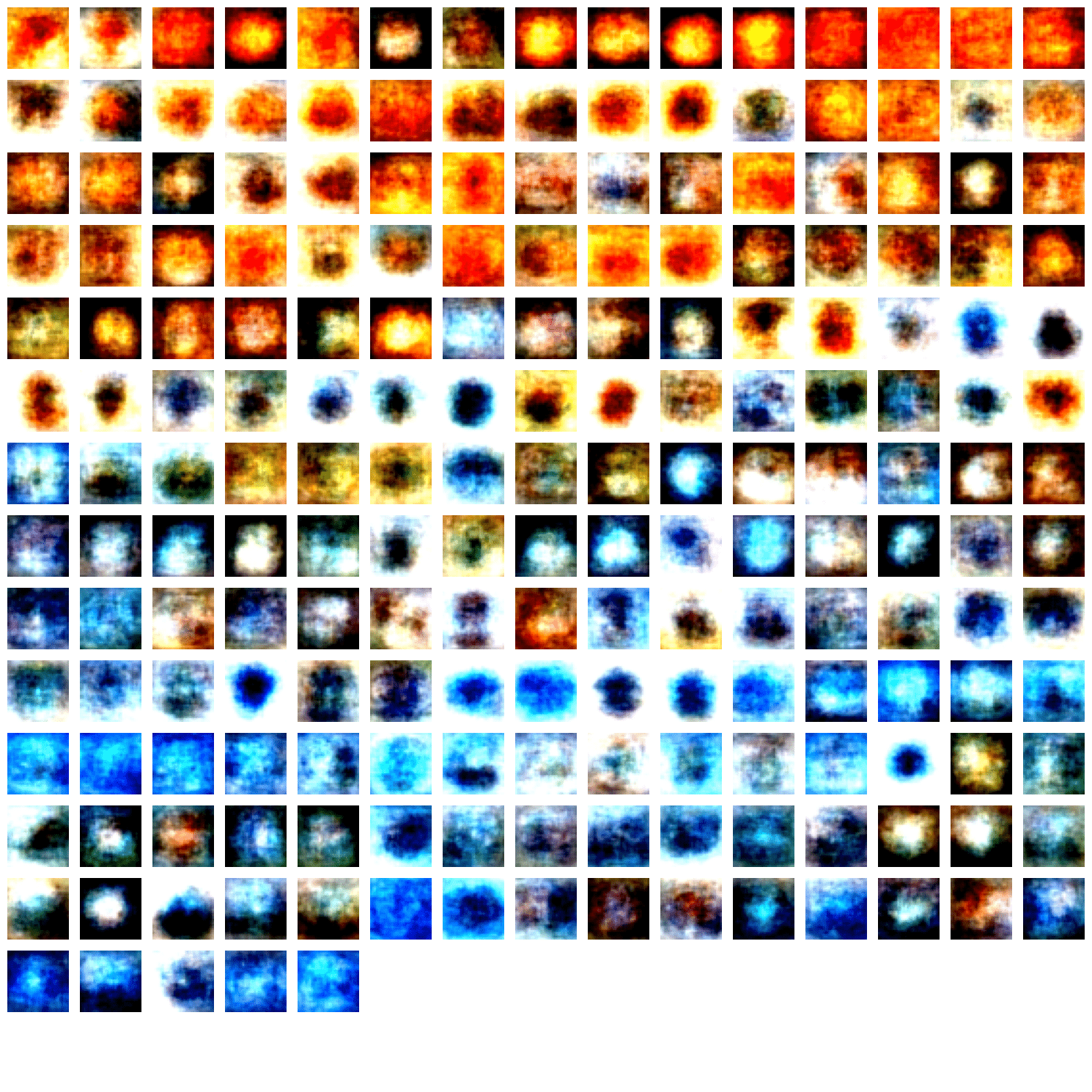}
    \caption{200 Subject-1 ICA reconstructions ordered by hierarchical clustering on the predicted ICA latents, which nicely organizes from warm to cold in terms of their hue (note that the warm to cold is not explicitly defined, and each subject is sorted by their own predicted ICA latents). The top and bottom 70 images are used for the ``red" and ``blue" group respectively}
    \label{fig:ica_ordering}
\end{figure}

\begin{figure}
    \centering
    \includegraphics[width=1\linewidth]{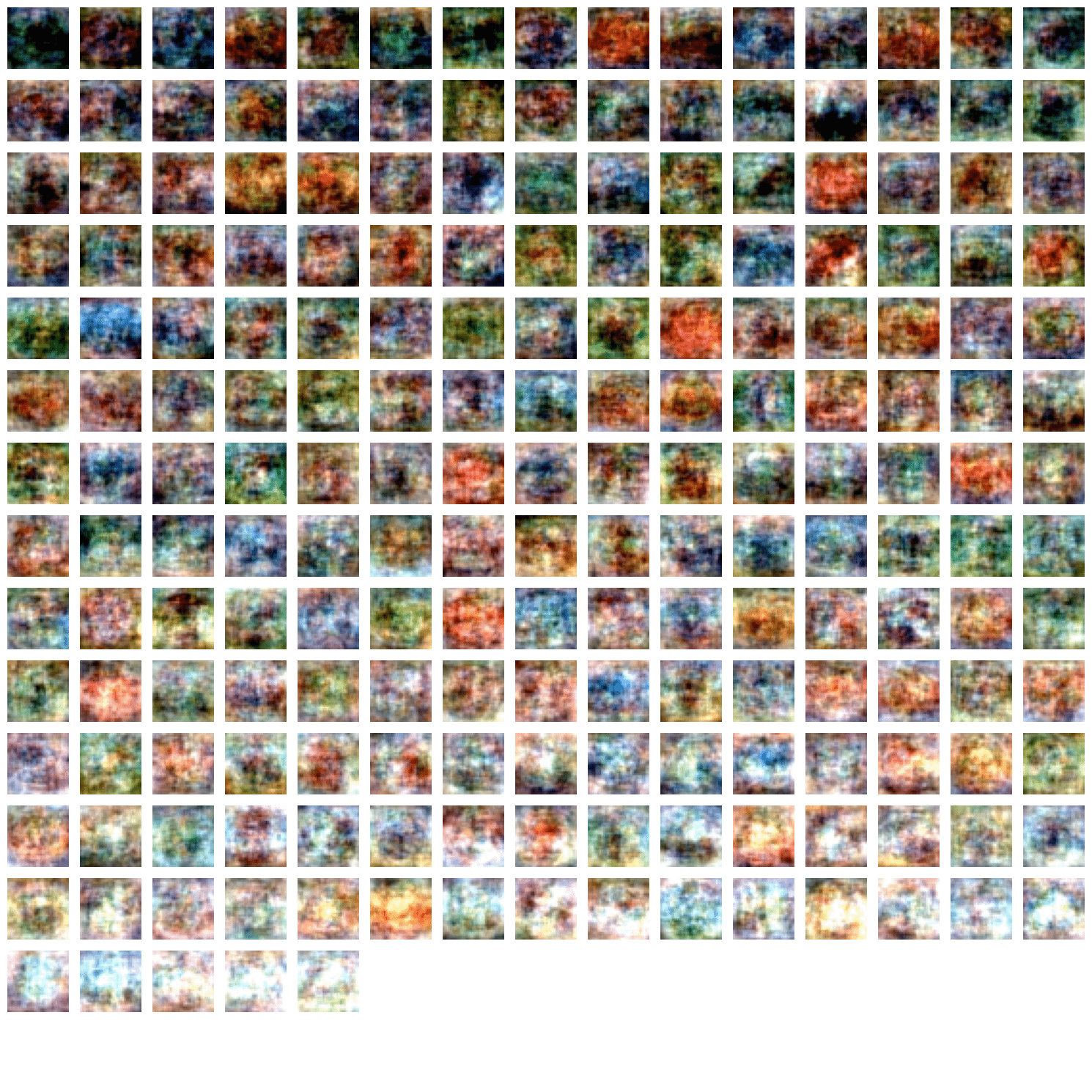}
    \caption{200 Subject-1 PCA reconstructions ordered by their luminance, increasing from left to right, and from top to bottom. Here we show an example ordering from  (each subject is sorted by their own reconstructions). The top and bottom 70 images are used for the ``dark" and ``bright" group respectively}
    \label{fig:pca_ordering}
\end{figure}

\begin{figure}
    \centering
    \includegraphics[width=1\linewidth]{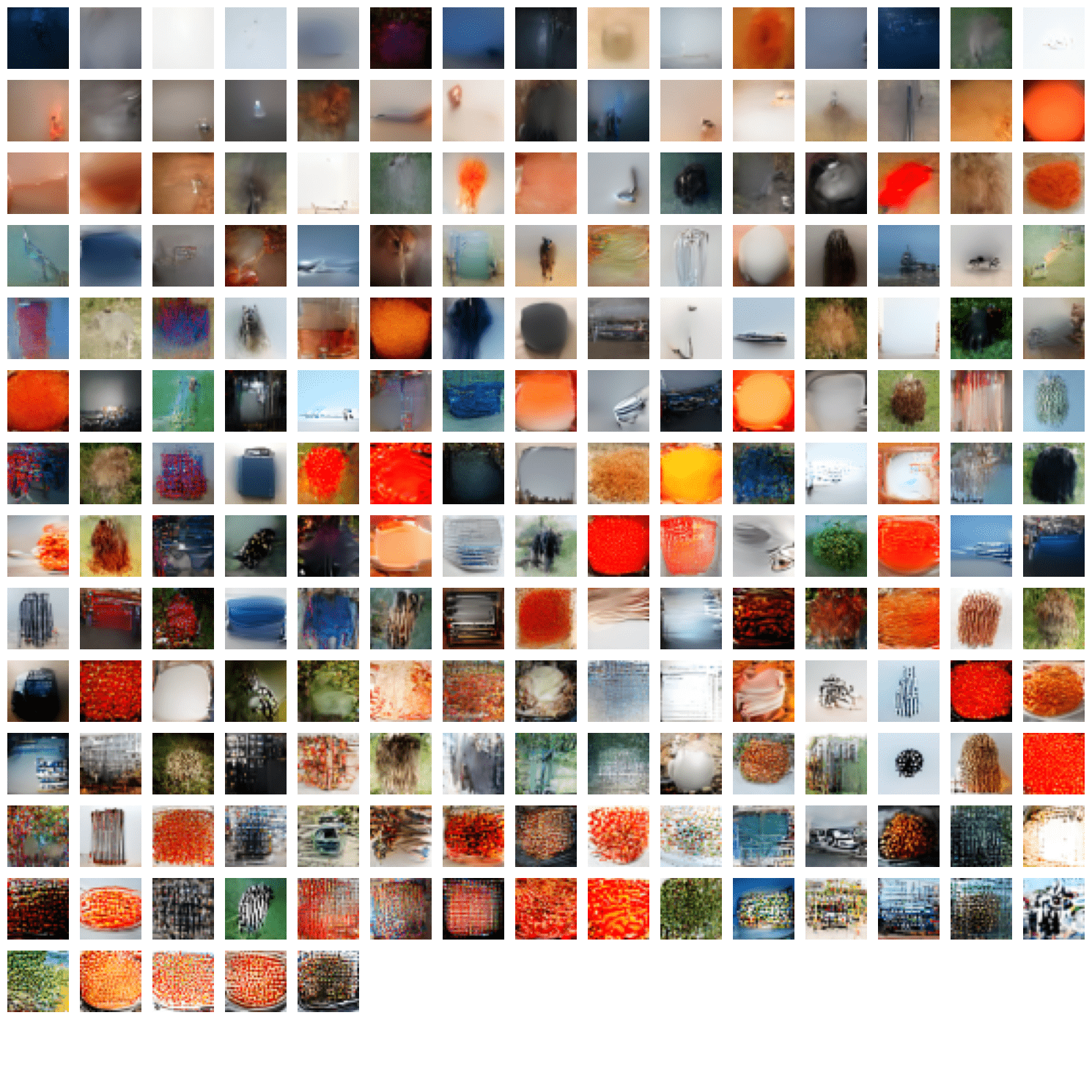}
    \caption{200 Subject-1 VDVAE reconstructions ordered by energy of the 2D FFT, increasing from left to right, and from top to bottom. Here we show an example ordering from  (each subject is sorted by their own reconstructions). The top and bottom 70 images are used for the ``smooth" and ``textured" group respectively}
    \label{fig:vdvae_ordering}
\end{figure}

The ordering for the PCA patterns, ICA patterns, and VDVAE patterns are each done slightly differently. The ordering for ICA is derived by hierarchical clustering on the predicted ICA latents, which automatically order them from red to blue (see Fig. \ref{fig:ica_ordering}), and the top and bottom 70 images are averaged into the ``red" and ``blue" group. The PCA patterns are sorted by the luminance (brightness level) of the reconstructions (see Fig. \ref{fig:pca_ordering}), and the top and bottom 70 images are averaged into ``dark" and ``bright" groups. The VDVAE patterns are sorted by the spatial energy (broadband power of the 2D FFT) which visually corresponds to smooth vs. textured (see Fig. \ref{fig:vdvae_ordering}), and top and bottom 70 images are averaged into the ``smooth" and ``textured groups". 

\section{Extended Results}
\begin{figure}
    \centering
    \includegraphics[width=0.6\linewidth]{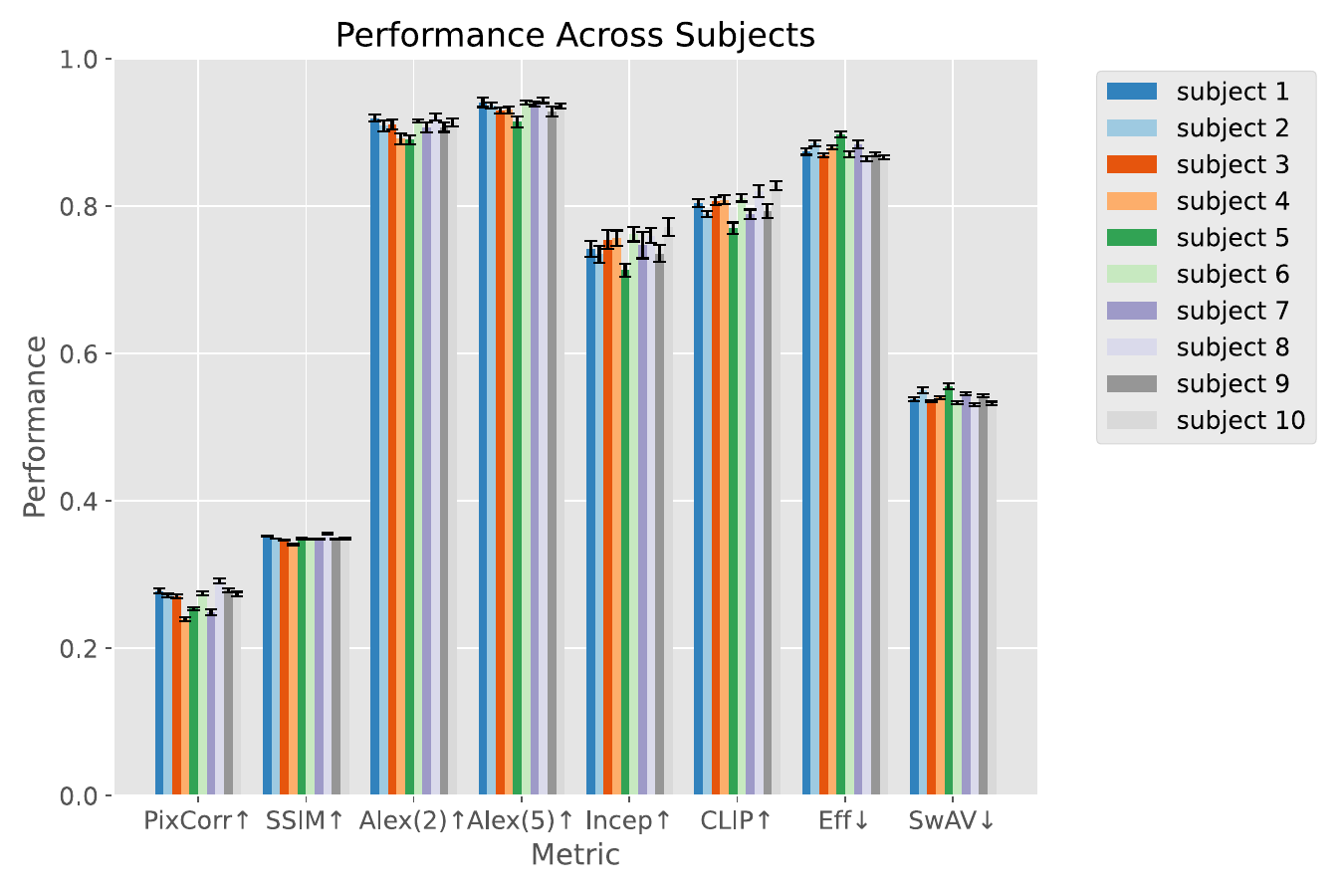}
    \caption{Performance across 10 subjects. It is computed by reconstructing with Versatile Diffusion using 7 different random seeds. For each subject, the final performance is the average across the 7 runs. The standard deviation across the 7 runs for each subject is represented by the error bars.}
    \label{fig:enter-label}
\end{figure}

\begin{figure}
    \centering
    \includegraphics[width=0.6\linewidth]{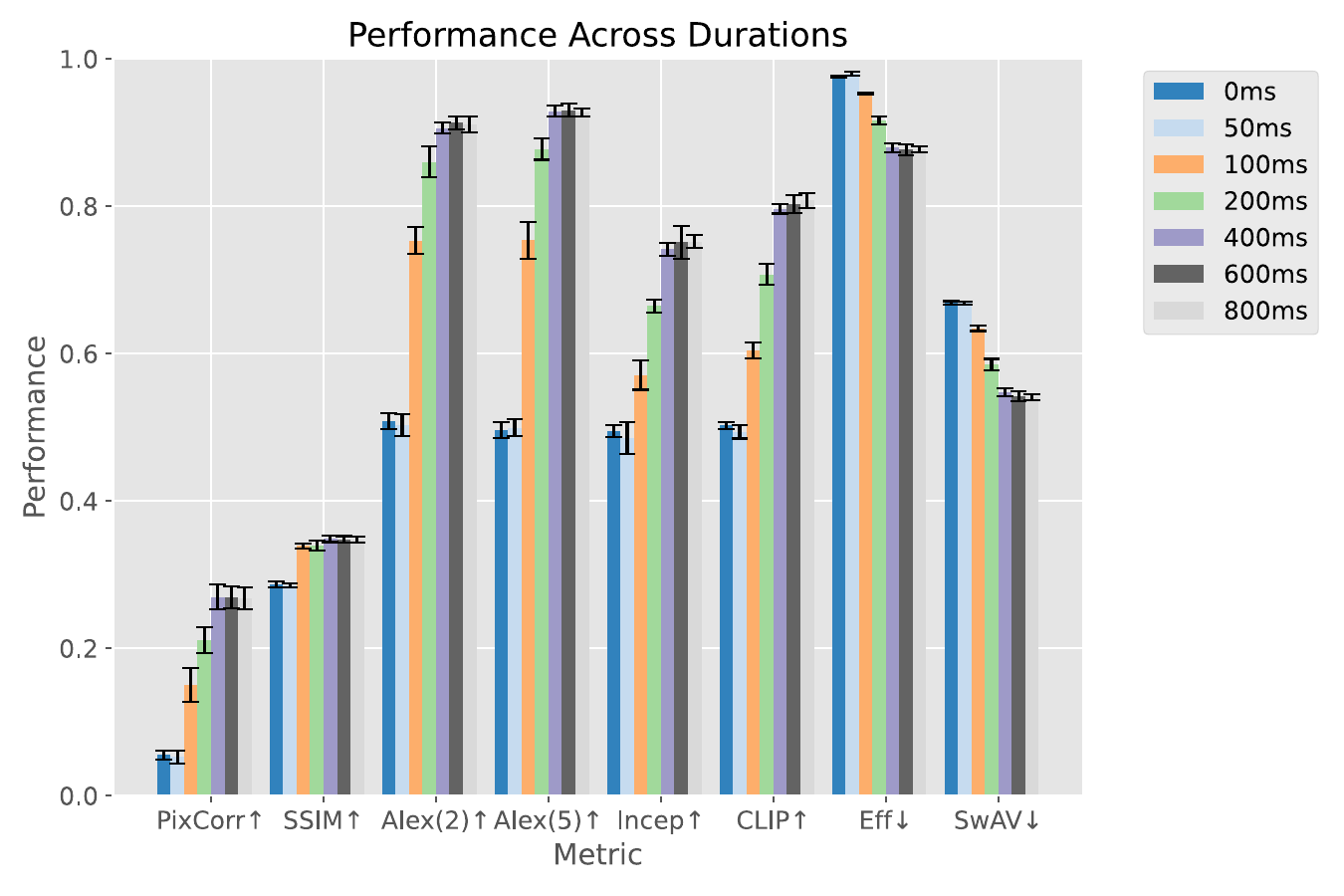}
    \caption{Performance across durations for Subjects 1 through 4. It is computed by models trained on each subject's 4-trial-averaged training data, and tested on their corresponding 80-trial-averaged test data. The ``first 200ms'', ``first 400ms'', ``first 600ms'' and ``first 800ms'' models use those corresponding time ranges after the onset of the stimulus. The 0ms performance, which should correspond to chance level, is computed by passing the 200ms before the onset of the stimulus onto the trained ``first 200ms'' model. The bars heights and the error bars represent the mean and standard deviation across the 4 subjects.}
    \label{fig:enter-label}
\end{figure}

\begin{figure}
    \centering
    \includegraphics[width=0.8\linewidth]{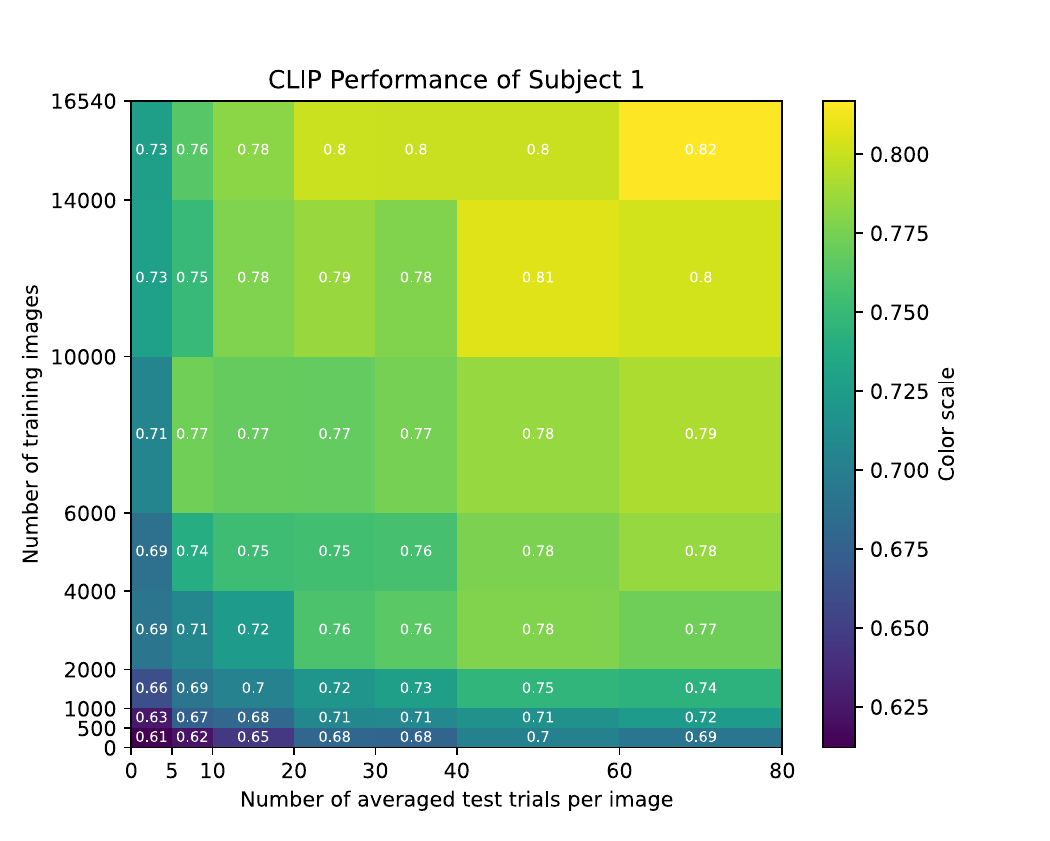}
    \caption{CLIP performance across training sizes and number of test trial averages shown for Subject 1 with random seed 0 used for reconstructions. It is computed by gradually increasing the the number of training images and the number of averages in the test samples. The y-axis shows gradual increase of the number of training images, and the x-axis shows gradual increase of the number of trial averages for each of the test images. Performance varies smoothly as a function of both training images and test trials.}
    \label{fig:enter-label}
\end{figure}

\begin{figure*}
    \centering
    \includegraphics[width=1\linewidth]{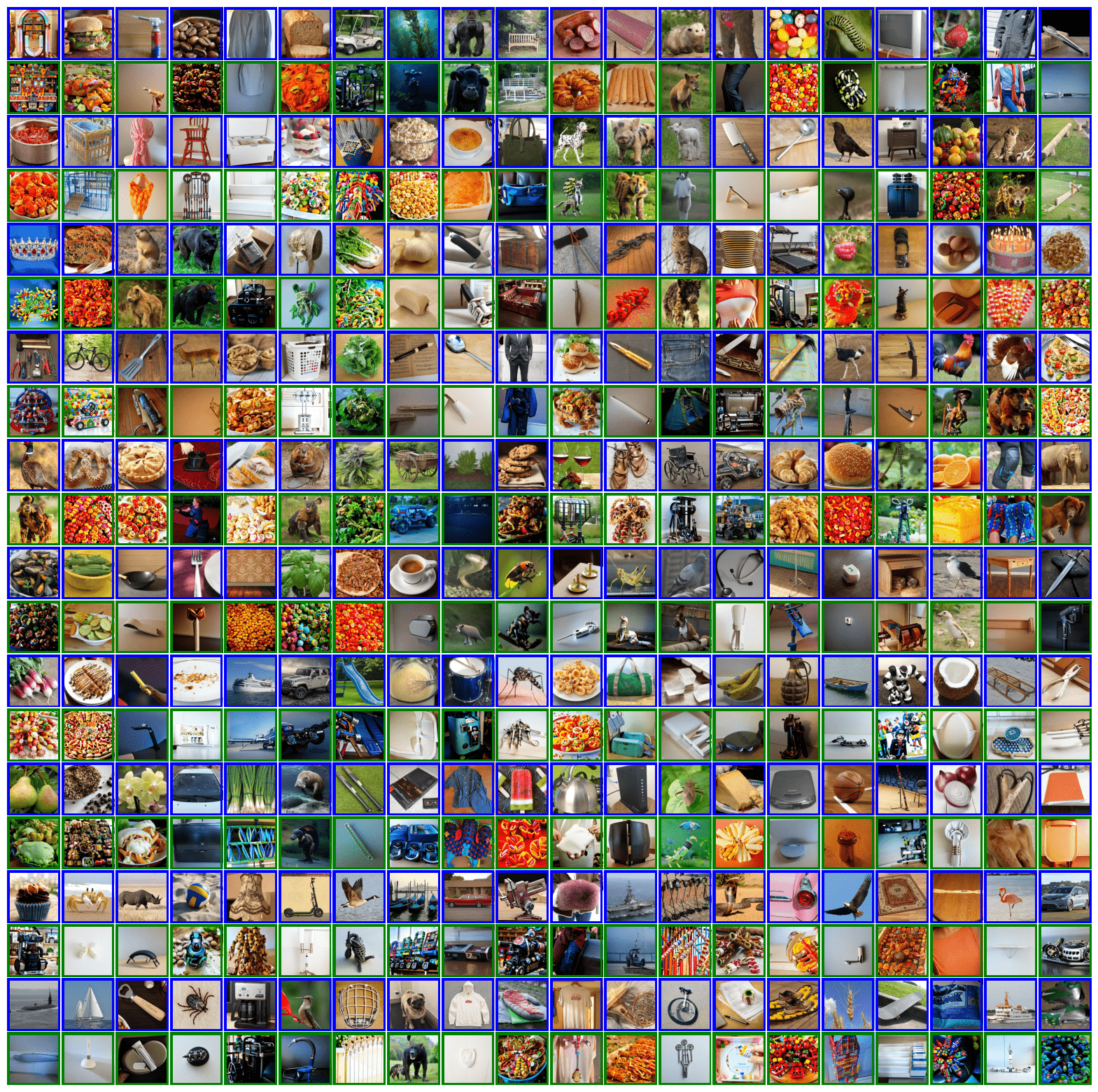}
    \caption{Full reconstructions (Subject 1) using the Versatile Diffusion reconstruction pipeline.}
    \label{fig:recon_plot_ordered_by_performance}
\end{figure*}
\begin{figure*}
    \centering
    \includegraphics[width=1\linewidth]{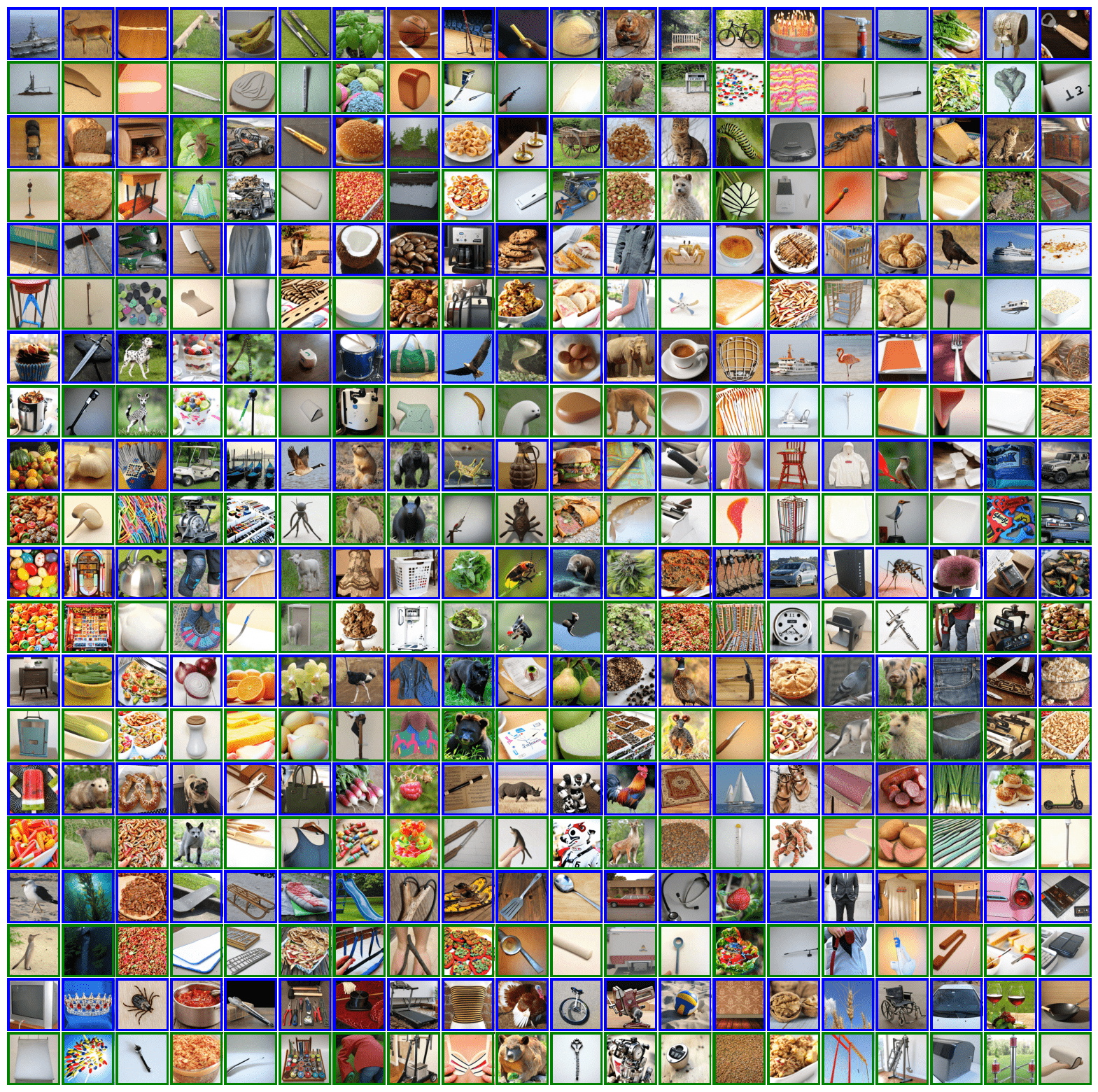}
    \caption{Full reconstructions (Subject 1) using the unCLIP reconstruction pipeline. }
    \label{fig:recon_unCLIP_full}
\end{figure*}
\begin{figure*}
    \centering
    \includegraphics[width=1\linewidth]{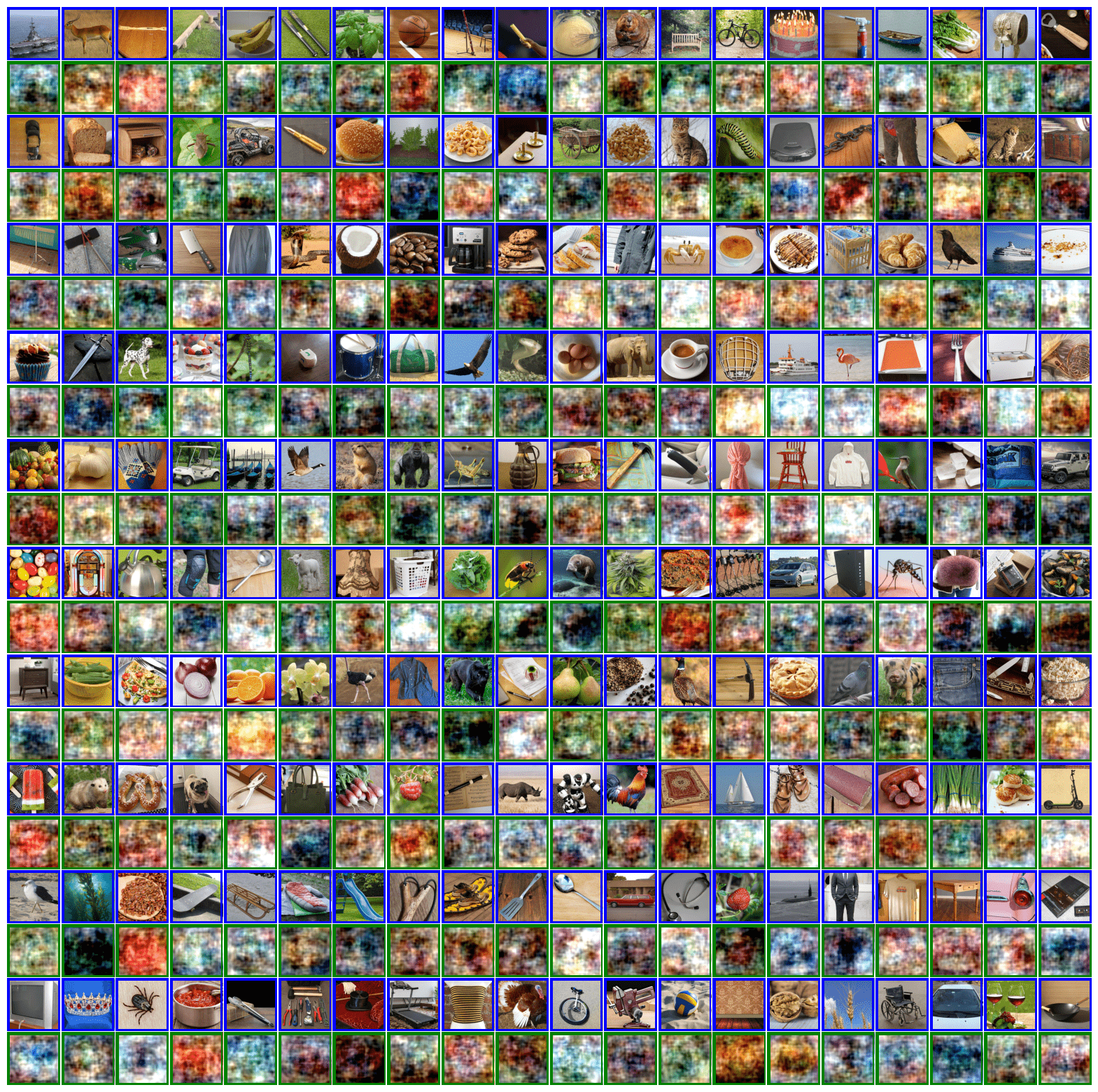}
    \caption{Full PCA reconstructions for Subject 1}
    \label{fig:PCA_recon_full}
\end{figure*}
\begin{figure*}
    \centering
    \includegraphics[width=1\linewidth]{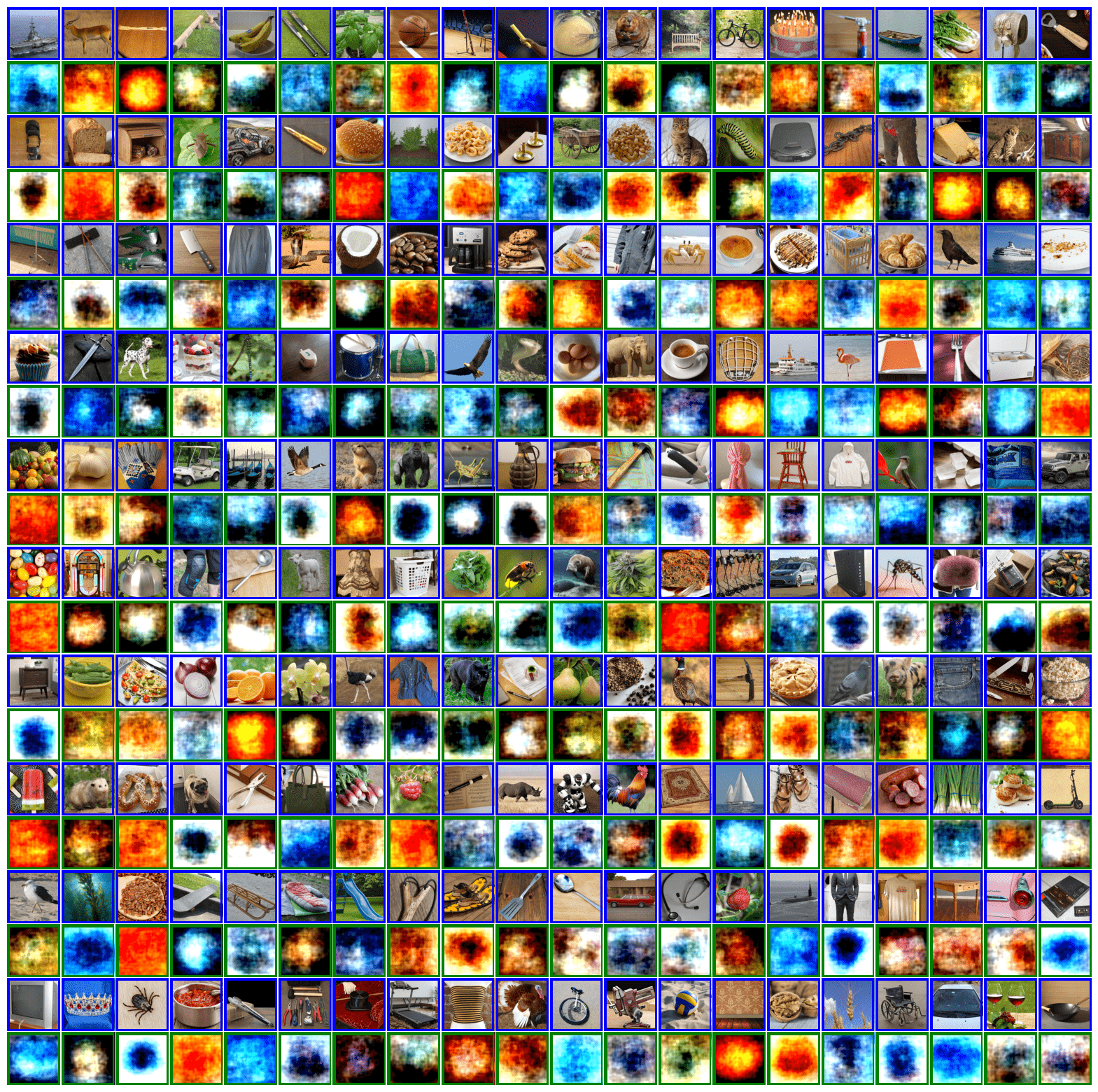}
    \caption{Full ICA reconstructions for Subject 1}
    \label{fig:ICA_recon_full}
\end{figure*}
\begin{figure*}
    \centering
    \includegraphics[width=1\linewidth]{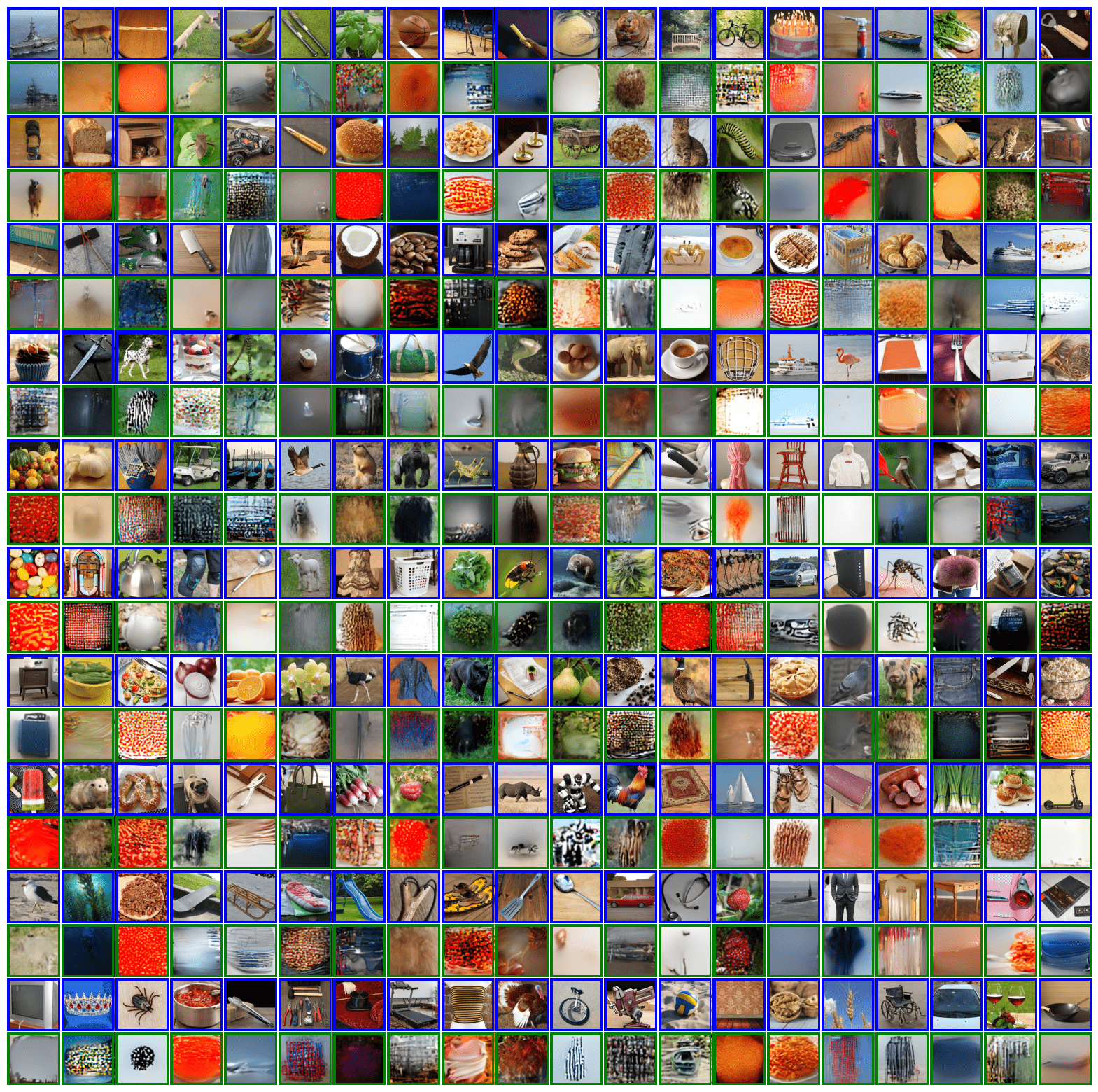}
    \caption{Full VDVAE reconstructions for Subject 1}
    \label{fig:recon_vdvae}
\end{figure*}

\begin{figure*}
    \centering
    \includegraphics[width=1\linewidth]{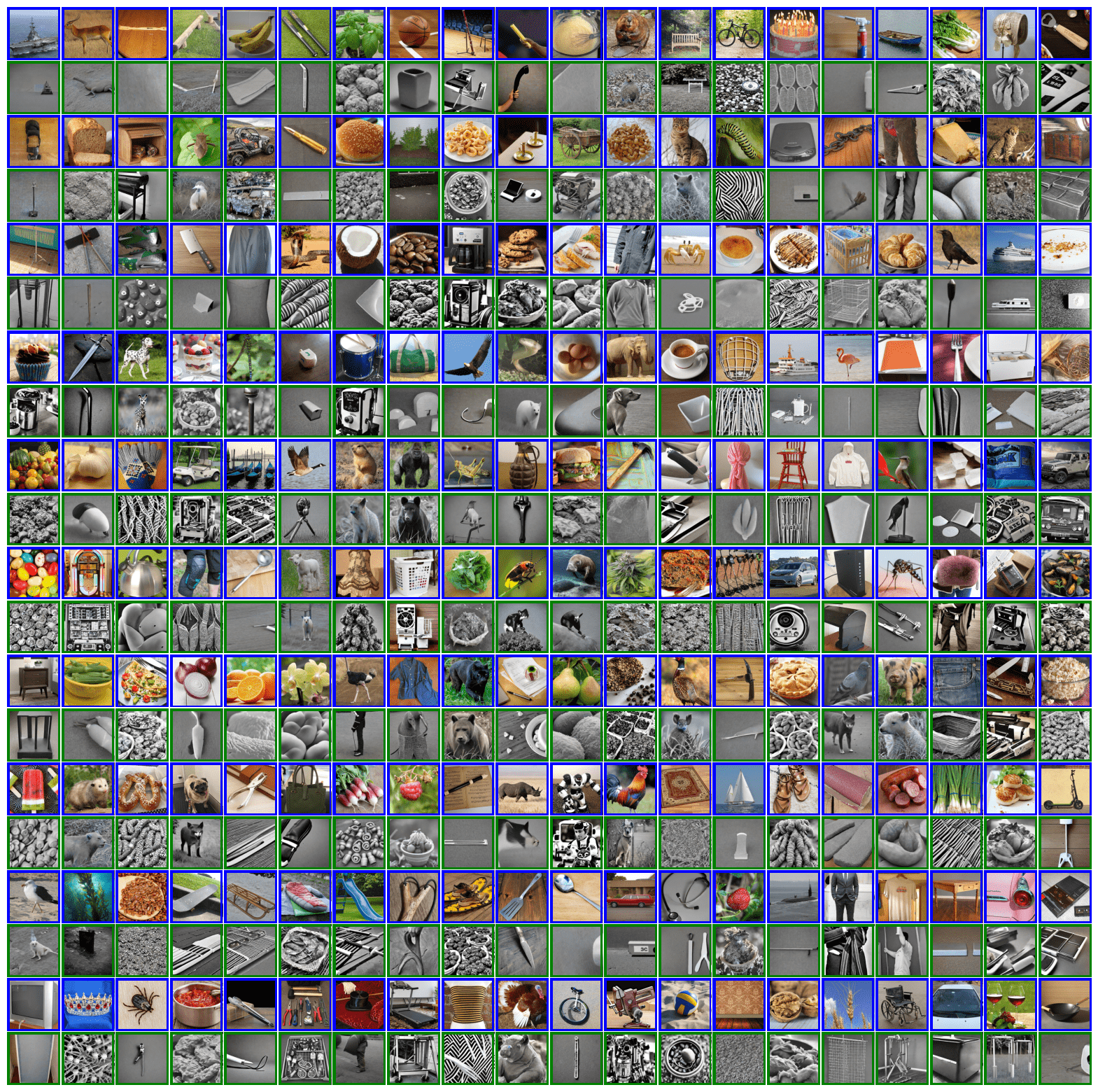}
    \caption{Full grayscale reconstructions (Subject 1) using the unCLIP reconstruction pipeline. The training images are converted into grayscale before being encoded into CLIP latnets. The rest of the pipeline remains the same.}
    \label{fig:recon_grayscale}
\end{figure*}

\begin{figure}
    \centering
    \includegraphics[width=1\linewidth]{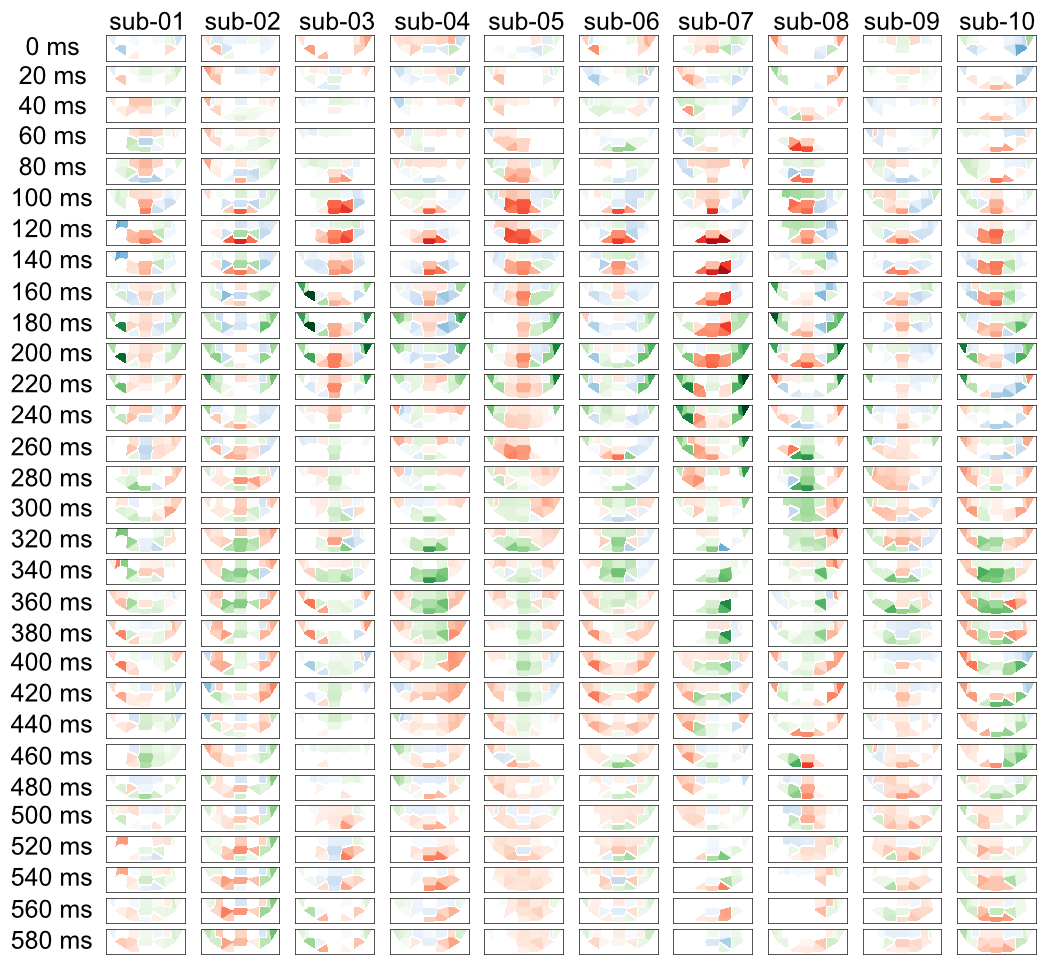}
    \caption{Whitened EEG of the 10 subjects}
    \label{fig:whitened}
\end{figure}
\begin{figure}
    \centering
    \includegraphics[width=1\linewidth]{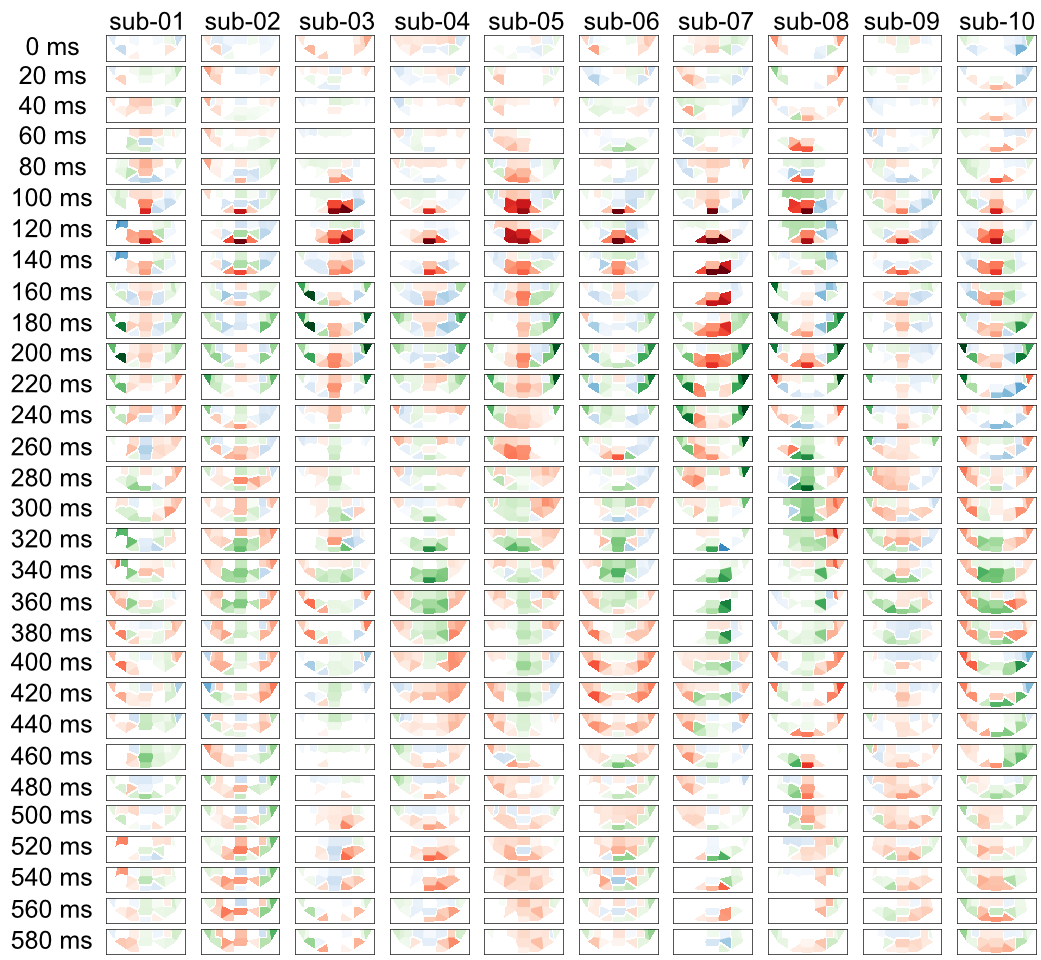}
    \caption{Mean-subtracted EEG of the 10 subjects}
    \label{fig:mean-subtracted}
\end{figure}

\begin{figure}
    \centering
    \includegraphics[width=1\linewidth]{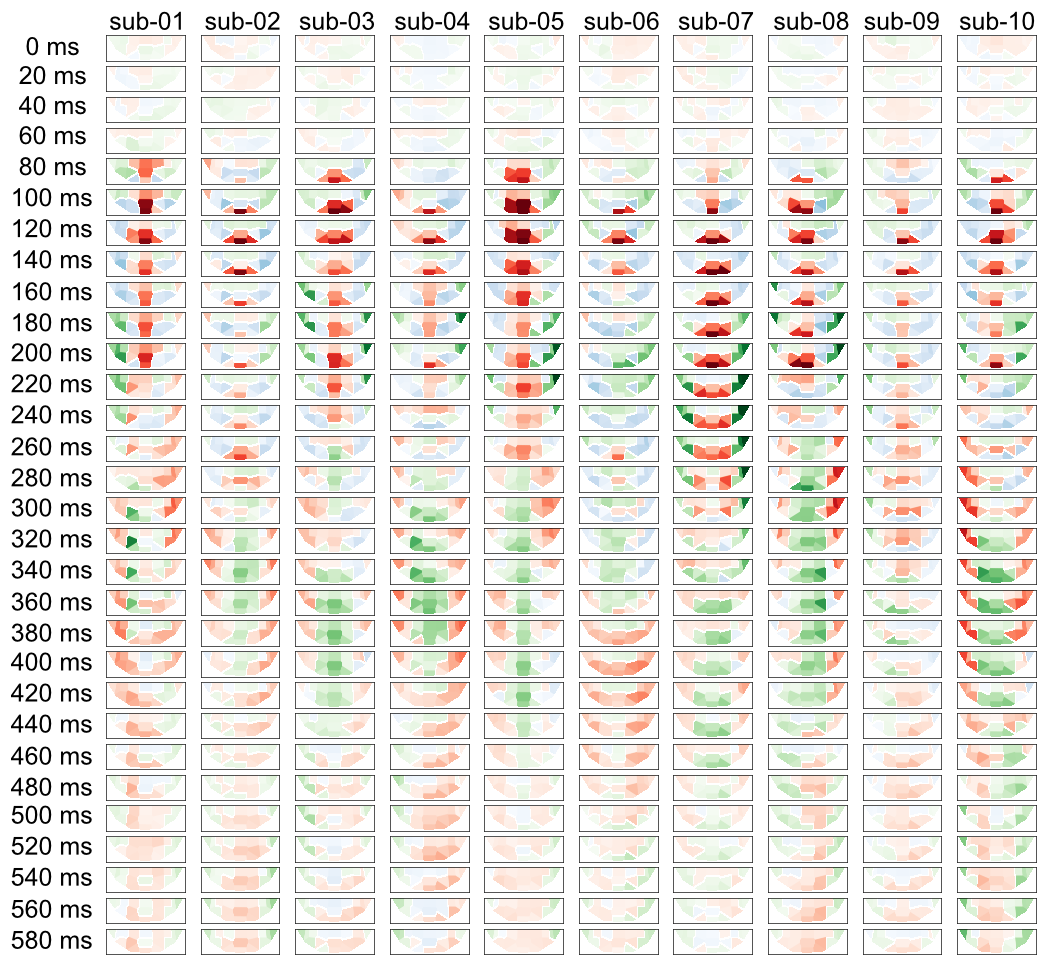}
    \caption{EEG patterns of the 10 subjects. 
    }
    \label{fig:indiv_eeg_patterns}
\end{figure}

\begin{figure}
    \centering
    \includegraphics[width=1\linewidth]{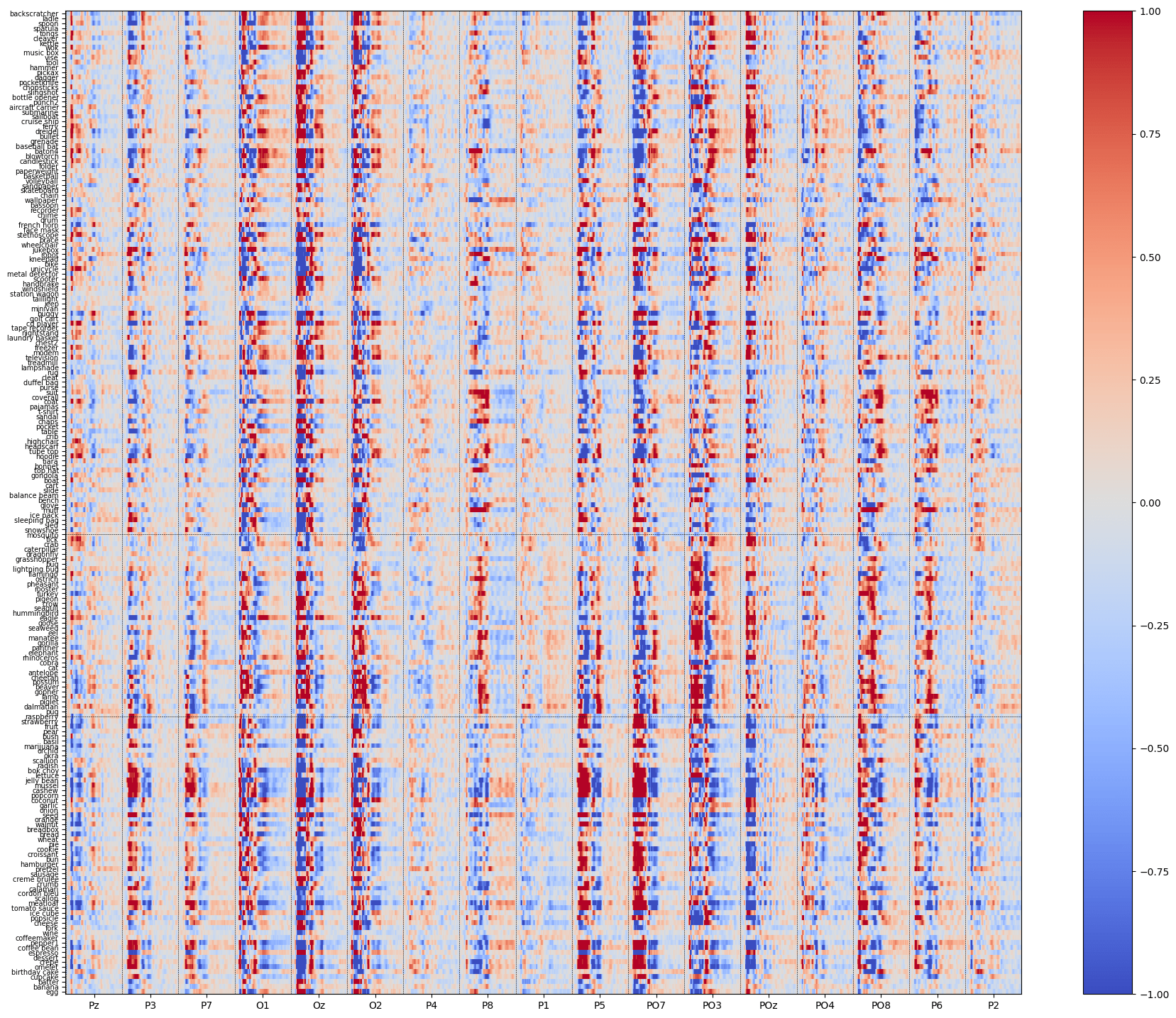}
    \caption{The full ``EEG pattern" of a single subject. Each row represents the EEG pattern of one of the 200 test images; the 2 horizontal dashed lines divide them into 3 general categories: food at the bottom, animals in the middle, and everything else at the top. The precise ordering was determined by hierarchical clustering of the CLIP representations of the images (not using EEG activity). Each column (between vertical black lines) represents an EEG channel; within each column, the smaller columns going from left to right are the time bins going from 0 to 800ms.  Note the consistency in the patterns within the food and animal categories reflecting similar brain activity underlying perception of these objects.}
    \label{fig:eeg_pattern_full}
\end{figure}
\begin{figure}
    \centering
    \includegraphics[width=1\linewidth]{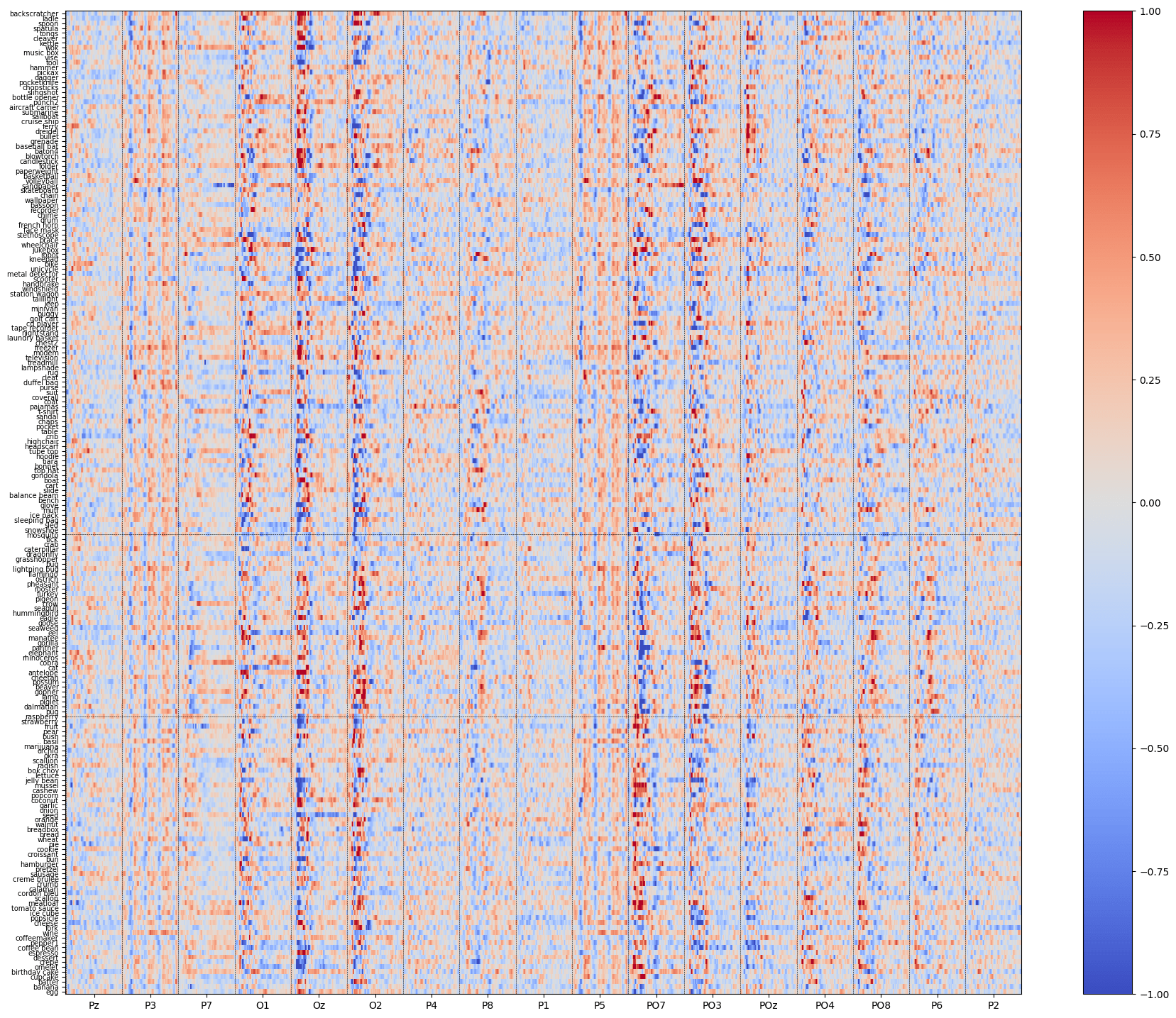}
    \caption{The real EEG of the same subject. Note that the differentiating features look less pronounced.}
    \label{fig:eeg_whitened}
\end{figure}

\begin{figure}
    \centering
    \includegraphics[width=0.4\linewidth]{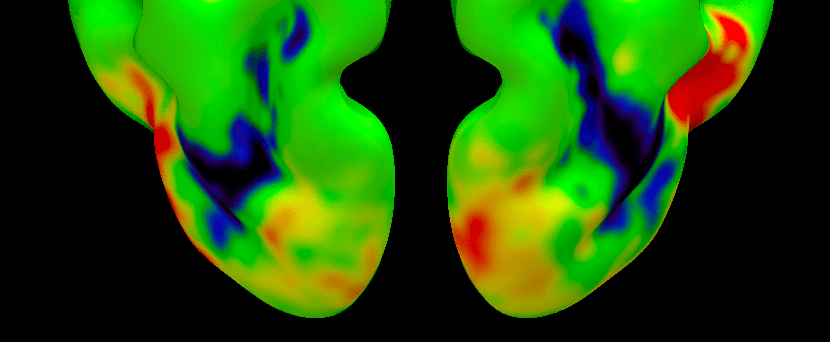}
    \caption{CLIP-fMRI pattern of the ``closeup human (faces)" category}
    \label{fig:human-closeup}
\end{figure}
\begin{figure}
    \centering
    \includegraphics[width=0.4\linewidth]{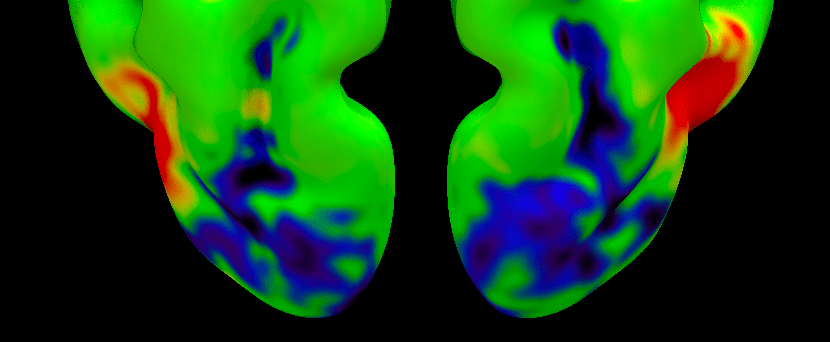}
    \caption{CLIP-fMRI pattern of the ``human from a distance" category}
    \label{fig:human-distant}
\end{figure}
\begin{figure}
    \centering
    \includegraphics[width=0.4\linewidth]{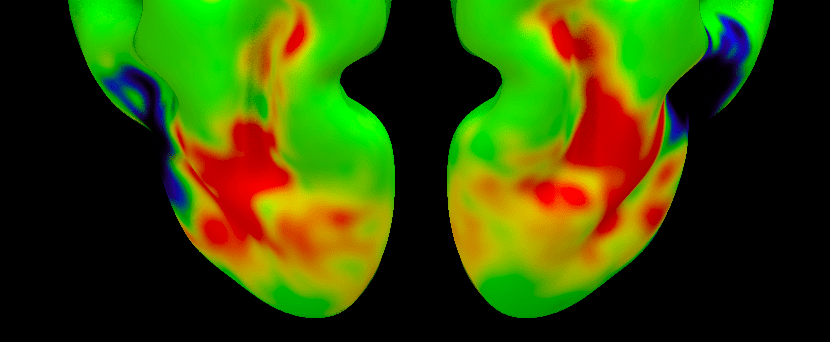}
    \caption{CLIP-fMRI pattern of the ``room interiors" category}
    \label{fig:interiors}
\end{figure}
\begin{figure}
    \centering
    \includegraphics[width=0.4\linewidth]{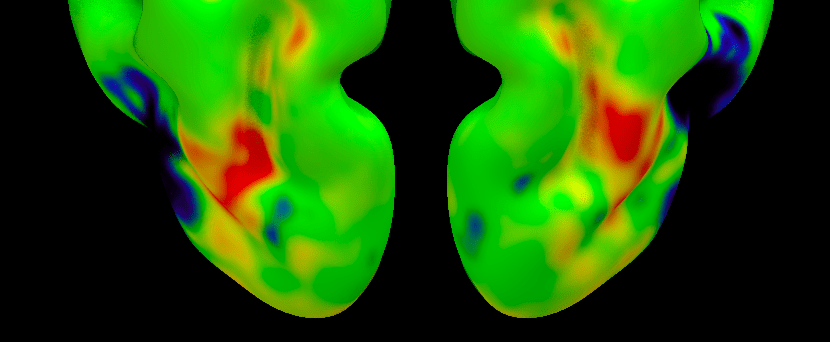}
    \caption{CLIP-fMRI pattern of the ``urban scenes" category}
    \label{fig:urban}
\end{figure}
\begin{figure*}
    \centering
    \includegraphics[width=0.9\linewidth]{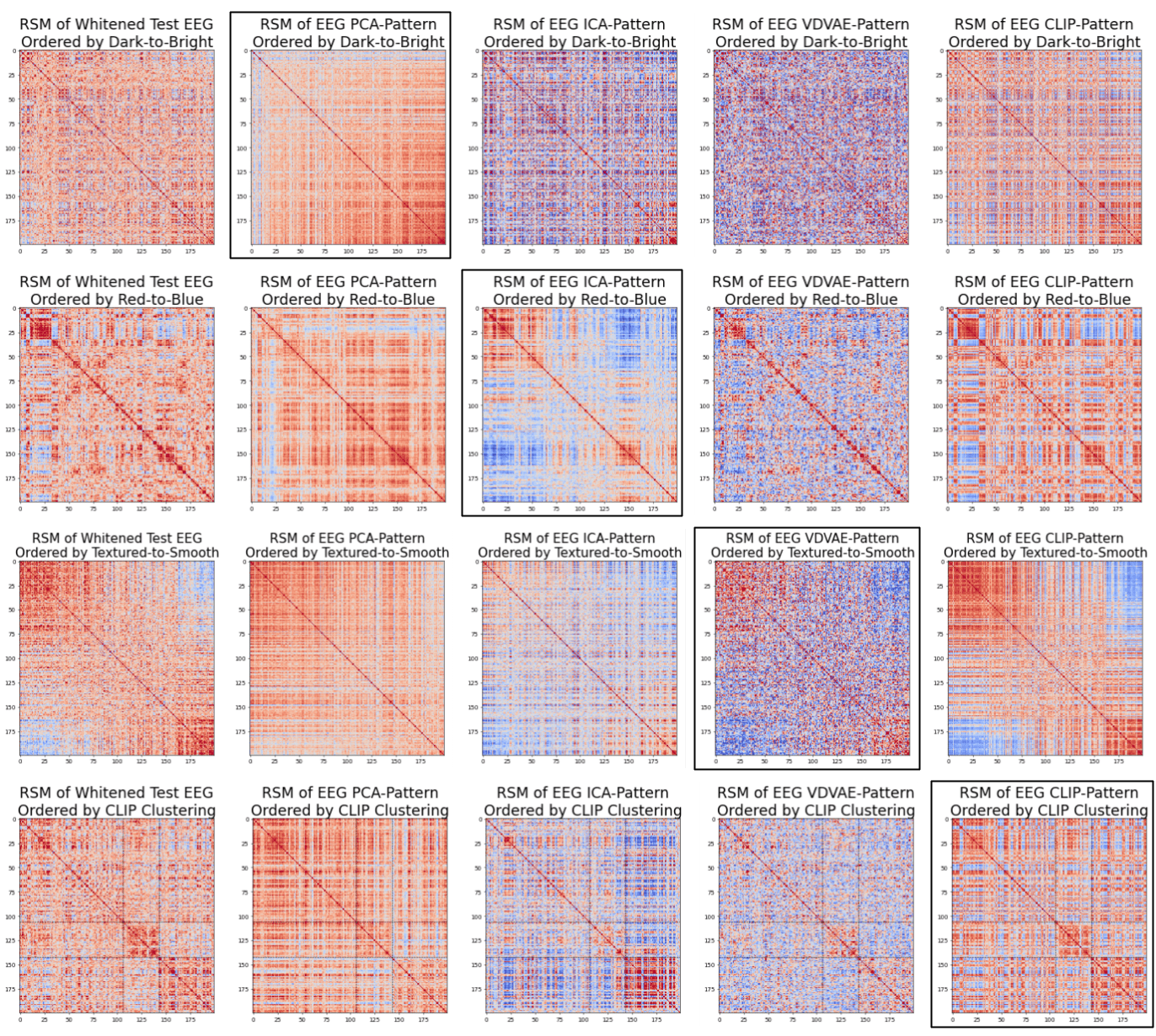}
    \caption{RSMs of Various Latent Spaces (Subject 1). Here, we show that the EEG data contains representational structure with respect to different low- and mid-level visual features of the stimulus, which is made evident when the RSM of the whitened EEG is ordered by luminance, color, and texture (left column). Similarly, each image latent space reliably encodes these visual features to variable degrees of selection. For example, PCA and ICA show structure for color and luminance, while VDVAE appears to select for the spatial frequency of the image. Finally, the EEG patterns generated from a given latent space (outlined subplots) exhibit representational structure for the visual feature(s) for which that latent space selects. 
We argue that, because CLIP encodes for these low- (color and luminance), mid- (texture), and high-level (semantic) features of visual stimuli, which are also encoded in EEG, a linear mapping is sufficient for preserving information between the two representational spaces.}
    \label{fig:RSM_clustering}
\end{figure*}

\end{document}